\documentclass[onecolumn,peerreview]{IEEEtran}
\ifCLASSINFOpdf
\else
    \usepackage[dvips]{graphicx}
\fi
\usepackage{amsmath,amssymb}
\newcommand{\bysame}{%
    \leavevmode\hbox to 3em{\hrulefill}\,}

\begin{document}
\title{A Further Note on an Innovations Approach to Viterbi Decoding of Convolutional Codes}

\author{Masato~Tajima,~\IEEEmembership{Senior~Member,~IEEE}
        %John~Doe,~\IEEEmembership{Fellow,~OSA,}
        %and~Jane~Doe,~\IEEEmembership{Life~Fellow,~IEEE}% <-this % stops a space
\thanks{M. Tajima is with University of Toyama, 3190 Gofuku,
Toyama 930-8555, Japan (e-mail: masatotjm@kind.ocn.ne.jp).}% <-this % stops a space
}


\markboth{Journal of \LaTeX\ Class Files,~Vol.~14, No.~8, August~2015}%
{Tajima \MakeLowercase{\textit{}}: A Further Note on an Innovations Approach to Viterbi Decoding of Convolutional Codes}



\maketitle

\begin{abstract}
In this paper, we show that the soft-decision input to the main decoder in an SST Viterbi decoder is regarded as the innovation as well from the viewpoint of mutual information and mean-square error. It is assumed that a code sequence is transmitted symbol by symbol over an AWGN channel using BPSK modulation. Then we can consider the signal model, where the signal is composed of the signal-to-noise ratio (SNR) and the equiprobable binary input. By assuming that the soft-decision input to the main decoder is the innovation, we show that the minimum mean-square error (MMSE) in estimating the binary input is expressed in terms of the distribution of the encoded block for the main decoder. It is shown that the obtained MMSE satisfies indirectly the known relation between the mutual information and the MMSE in Gaussian channels. Thus the derived MMSE is justified, which in turn implies that the soft-decision input to the main decoder can be regarded as the innovation. Moreover, we see that the input-output mutual information is connected with the distribution of the encoded block for the main decoder.
\end{abstract}


% \fi
\IEEEpeerreviewmaketitle

\section{Introduction}
Consider an SST Viterbi decoder~\cite{kubo 93,taji 032} which consists of a pre-decoder and a main decoder. In~\cite{taji 19}, by comparing with the results in the linear filtering theory~\cite{ari 77,jaz 70,kai 681,kai 682,kai 74}, we showed that the hard-decision input to the main decoder can be seen as the {\it innovation}~\cite{kai 681,kai 682,kai 98,kuni 76}. In the coding theory, the framework of the theory is determined by hard-decision data. In the previous paper~\cite[Definition 1]{taji 19}, we have given the definition of innovations for hard-decision data. We see that the definition applies also to soft-decision data. That is, if the associated hard-decision data is the innovation, then the original soft-decision data can be regarded as the innovation. Hence, we can consider the {\it soft-decision} input to the main decoder to be the innovation as well. In this paper, concerning this subject, we show that this is true from a different viewpoint. We use the innovations approach to least-squares estimation by Kailath~\cite{kai 681,kai 682} and the relationship between the mutual information and the mean-square error (e.g.,~\cite{guo 05}). 
\par
Consider the situation where we have observations of a signal in additive white Gaussian noise. (In this paper, it is assumed that the signal is composed of the {\it signal-to-noise ratio} (SNR) and the equiprobable binary input, where the latter is not dependent on the SNR. In the following, we call the equiprobable binary input simply the input.) Kailath~\cite{kai 681,kai 682} applied the innovations method to linear filtering/smoothing problems. In the {\it discrete-time} case~\cite{kai 681,kai 74}, he showed that the covariance matrix of the innovation is expressed as the sum of two covariance matrices, where one is corresponding to the estimation error of the signal and the other is corresponding to the observation noise. Then the mean-square error in estimating the signal is obtained from the former covariance matrix by taking its trace. Based on the result of Kailath, we thought as follows: if the soft-decision input to the main decoder corresponds to the innovation, then the associated covariance matrix should have the above property. That is, the covariance matrix of the soft-decision input to the main decoder can be decomposed as the sum of two covariance matrices and hence by following the above, the mean-square error in estimating the signal will be obtained. We remark that in this context, the obtained mean-square error has to be justified using some method. For the purpose, we have noted the relation between the mutual information and the mean-square error.
\par
By the way, we derived the distribution of the input to the main decoder corresponding to a single code symbol in~\cite{taji 19}. However, we could not obtain the joint distribution of the input to the main decoder corresponding to a branch. After our paper~\cite{taji 19} was published, we have noticed that the distribution of the input to the main decoder corresponding to a single code symbol has a reasonable interpretation (see~\cite{taji 20}). Then using this fact, the joint distribution of the input to the main decoder corresponding to a branch has been derived. In that case, the associated covariance matrix can be calculated. Then our argument for obtaining the mean-square error is as follows: 
\begin{itemize}
\item [1)] Derive the joint distribution of the soft-decision input to the main decoder.
\item [2)] Calculate the associated covariance matrix using the derived joint distribution.
\item [3)] The covariance matrix of the estimation error of the signal is obtained by subtracting that of the observation noise from the matrix in 2). Moreover, by removing the SNR from the obtained covariance matrix, the covariance matrix of the estimation error of the input is derived.
\item [4)] The mean-square error in estimating the input is given by the trace (i.e., the sum of diagonal elements) of the last covariance matrix.
\end{itemize}
In this way, we show that the mean-square error in estimating the input is expressed in terms of the distribution of the encoded block for the main decoder. Since in an SST Viterbi decoder, the estimation error is decoded at the main decoder, the result is reasonable.
\par
On the other hand, we have to show the validity of the derived mean-square error. For the purpose, we note the relation between the mutual information and the mean-square error. Including these notions, the mutual information, the {\it likelihood ratio} (LR), and the mean-square error are central concerns in information theory, detection theory, and estimation theory. It is well known that the best estimate measured in mean-square sense (i.e., the least-squares estimate) is given by the conditional expectation~\cite{kai 98,kuni 76,oksen 98,will 91}. In this paper, the corresponding mean-square error is referred to as the {\it minimum mean-square error} (MMSE)~\cite{guo 05}. In this case, depending on the amount of observations $\{z(t)\}$ used for the estimation, the {\it causal} (filtering) MMSE and the {\it noncausal} (smoothing) MMSE are considered. When $\{z(\tau),~\tau \leq t\}$ is used for the estimation of the input $x(t)$, it corresponds to filtering, whereas when $\{z(\tau),~\tau \leq T\}$ is used for the estimation of $x(t)~(t<T)$, it corresponds to smoothing.
\par
From the late 1960's to the early 1970's, the relation between the LR and the mean-square error was actively discussed. Kailath~\cite{kai 69} showed that the LR is expressed in terms of the causal least-squares estimate. Moreover~\cite{kai 683}, he showed that the causal least-squares estimate can be obtained from the LR. Esposito~\cite{espo 68} also derived the relation between the LR and the noncausal estimator, which is closely related to~\cite{kai 683}. Subsequently, Duncan~\cite{dun 70} and Kadota et al.~\cite{kado 71} derived the relation between the mutual information and the causal mean-square error. Their works are regarded as extensions of the work of Gelfand and Yaglom (see~\cite{ari 77,dun 70}), who discussed for the first time the relation between the mutual information and the filtering error. Later, Guo et al.~\cite{guo 05} derived new formulas regarding the relation between the mutual information and the noncausal mean-square error. Furthermore, Zakai~\cite{zak 05} showed that the relation between the mutual information and the estimation error holds also in the abstract Wiener Space~\cite{wong 71,zak 05}. He applied not the Ito calculus~\cite{ito 76,oksen 98} but the Malliavin calculus~\cite{mall 76,zak 05}. We remark that the additive Gaussian noise channel is assumed for all above works.
\par
Among those works, we have noted the work of Guo et al.~\cite{guo 05}. It deals with the relation between the mutual information and the MMSE. Also, their signal model is equal to ours and hence it seems favorable for manipulating. We thought if the MMSE obtained using the above method satisfies the relations in~\cite[Section IV-A]{guo 05}, then the derived MMSE is justified, which in turn implies that the soft-decision input to the main decoder can be regarded as the innovation. The main text of the paper consists of the arguments stated above. By the way, it is shown that the MMSE is expressed in terms of the distribution of the encoded block for the main decoder. Then we see that the input-output mutual information is connected with the distribution of the encoded block for the main decoder. We think this is an extension of the relation between the mutual information and the MMSE to coding theory. The remainder of this paper is organized as follows.
\par
In Section II, based on the signal model in this paper, we derive the joint distribution of the input to the main decoder.
\par
In Section III, the associated covariance matrix is calculated using the derived joint distribution. Subsequently, by subtracting the covariance matrix of the observation noise from it and by removing the SNR from the resulting matrix, the covariance matrix of the estimation error of the input is obtained. The MMSE in estimating the input is given by the trace of the last matrix. In this case, since the diagonal elements of the target matrix have a correlation, we modify the obtained MMSE. Note that this modification is essentially important. (We remark that some results in Sections II and III have been given in~\cite{taji 20} with or without proofs. However, in order for the paper to be self-contained, the necessary materials are provided again with proofs in those sections.)
\par
In Section IV, the validity of the derived MMSE is discussed. The argument is based on the relation between the mutual information and the MMSE. More precisely, we discuss using the results of Guo et al.~\cite[Section IV-A]{guo 05}. We remark that the input in our signal model is not Gaussian. Hence the input-output mutual information cannot be obtained in a concrete form. We have only inequalities and approximate expressions. For that reason, we carry out numerical calculations using concrete convolutional codes. In this case, in order to clarify the difference between causal estimation (filtering) and noncausal one (smoothing), we take QLI codes~\cite{mass 71} with different constraint-lengths. Then the MMSE's are calculated by regarding these QLI codes as general codes on one side and by regarding them as inherent QLI codes on the other. The obtained results are compared and carefully examined. Moreover, through the argument, we see that the input-output mutual information is connected with the distribution of the encoded block for the main decoder.
\par
In Section V, we give several important comments regarding the discussions in Section IV.
\par
Finally, in Section IV, we conclude with the main points of the paper and with problems to be further discussed.
\par
Let us close this section by introducing the basic notions needed for this paper. Notations in this paper are same as those in~\cite{taji 19} in principle. We always assume that the underlying field is $\mbox{GF}(2)$. Let $G(D)$ be a generator matrix for an $(n_0, k_0)$ convolutional code, where $G(D)$ is assumed to be {\it minimal}~\cite{forn 70}. A corresponding check matrix $H(D)$ is also assumed to be minimal. Hence they have the same constraint length, denoted $\nu$. Denote by $\mbox{\boldmath $i$}=\{\mbox{\boldmath $i$}_k\}$ and $\mbox{\boldmath $y$}=\{\mbox{\boldmath $y$}_k\}$ an information sequence and the corresponding code sequence, respectively, where $\mbox{\boldmath $i$}_k=(i_k^{(1)}, \cdots, i_k^{(k_0)})$ is the information block at $t=k$ and $\mbox{\boldmath $y$}_k=(y_k^{(1)}, \cdots, y_k^{(n_0)})$ is the encoded block at $t=k$. In this paper, it is assumed that a code sequence {\boldmath $y$} is transmitted symbol by symbol over a memoryless AWGN channel using BPSK modulation. Let $\mbox{\boldmath $z$}=\{\mbox{\boldmath $z$}_k\}$ be a received sequence, where $\mbox{\boldmath $z$}_k=(z_k^{(1)}, \cdots, z_k^{(n_0)})$ is the received block at $t=k$. Each component $z_j$ of {\boldmath $z$} is modeled as
\begin{eqnarray}
z_j &=& x_j\sqrt{2E_s/N_0}+w_j \\
&=& cx_j+w_j ,
\end{eqnarray}
where $c\stackrel{\triangle}{=}\sqrt{2E_s/N_0}$. Here, $x_j$ takes $\pm 1$ depending on whether the code symbol $y_j$ is $0$ or $1$. That is, $x_j$ is the equiprobable binary input. (We call it simply the input.) $E_s$ and $N_0$ denote the energy per channel symbol and the single-sided noise spectral density, respectively. (Let $E_b$ be the energy per information bit. Then the relationship between $E_b$ and $E_s$ is defined by $E_s=RE_b$, where $R$ is the code rate.) Also, $w_j$ is a zero-mean unit variance Gaussian random variable with probability density function
\begin{equation}
q(y)=\frac{1}{\sqrt{2\pi}}e^{-\frac{y^2}{2}} .
\end{equation}
Each $w_j$ is independent of all others. The hard-decision (denoted ``$^h$'') data of $z_j$ is defined by
\begin{equation}
z_j^h\stackrel{\triangle}{=}\left\{
\begin{array}{rl}
0,& \quad z_j \geq 0 \\
1,& \quad z_j < 0 .
\end{array} \right.
\end{equation}
In this case, the channel error probability (denoted $\epsilon$) is given by
\begin{equation}
\epsilon=\frac{1}{\sqrt{2\pi}} \int_{\sqrt{2E_s/N_0}}^{\infty}e^{-\frac{y^2}{2}}dy \stackrel{\triangle}{=}Q\bigl(\sqrt{2E_s/N_0}\bigr) .
\end{equation}
Note that the above signal model can be seen as the block at $t=k$. In that case, we can rewrite as
\begin{equation}
\mbox{\boldmath $z$}_k=c\mbox{\boldmath $x$}_k+\mbox{\boldmath $w$}_k ,
\end{equation}
where $\mbox{\boldmath $x$}_k=(x_k^{(1)}, \cdots, x_k^{(n_0)})$, $\mbox{\boldmath $w$}_k=(w_k^{(1)}, \cdots, w_k^{(n_0)})$, and $\mbox{\boldmath $z$}_k=(z_k^{(1)}, \cdots, z_k^{(n_0)})$.
\par
In this paper, we consider an SST Viterbi decoder~\cite{kubo 93,taji 032,taji 19} which consists of a pre-decoder and a main decoder. In the case of a general convolutional code, the inverse encoder $G^{-1}(D)$ is used as a pre-decoder. Let $\mbox{\boldmath $r$}_k=(r_k^{(1)}, \cdots, r_k^{(n_0)})$ be the soft-decision input to the main decoder. $r_k^{(l)}~(1 \leq l \leq n_0)$ is given by
\begin{equation}
r_k^{(l)}=\left\{
\begin{array}{rl}
\vert z_k^{(l)} \vert,& \quad r_k^{(l)h}=0 \\
- \vert z_k^{(l)} \vert,& \quad r_k^{(l)h}=1 .
\end{array} \right.
\end{equation}

\section{Joint Distribution of the Input to the Main Decoder}
The argument in this paper is based on the result of Kailath~\cite[Section III-B]{kai 681}. That is, by calculating the covariance matrix of the input to the main decoder, we derive the mean-square error in estimating the input. First we obtain the joint distribution of the input to the main decoder. In the following, $P(\,\cdot\,)$ and $E[\,\cdot\,]$ denote the probability and expectation, respectively.
\par
Let $\mbox{\boldmath $r$}_k=(r_k^{(1)}, \cdots, r_k^{(n_0)})$ be the input to the main decoder in an SST Viterbi decoder~\cite{taji 19}. In connection with the distribution of $r_k^{(l)}~(1 \leq l \leq n_0)$~\cite[Proposition 12]{taji 19}, $\alpha~(=\alpha_l)$ is defined by
\begin{equation}
\alpha\stackrel{\triangle}{=}P(e_k^{(l)}=0, r_k^{(l)h}=1)+P(e_k^{(l)}=1, r_k^{(l)h}=0) .
\end{equation}
We have noticed that this value has another interpretation. In fact, we have the following.
\newtheorem{lem}{Lemma}
\begin{lem}[\cite{taji 20}]
\begin{equation}
\alpha_l=P(v_k^{(l)}=1)~(1 \leq l \leq n_0)
\end{equation}
holds, where $\mbox{\boldmath $v$}_k=(v_k^{(1)}, \cdots, v_k^{(n_0)})$ is the encoded block for the main decoder.
\end{lem}
\begin{IEEEproof}
The hard-decision input to the main decoder is expressed as
\begin{displaymath}
\mbox{\boldmath $r$}_k^h=\mbox{\boldmath $u$}_kG+\mbox{\boldmath $e$}_k ,
\end{displaymath}
where $\mbox{\boldmath $u$}_k=\mbox{\boldmath $e$}_kG^{-1}$. Let $\mbox{\boldmath $v$}_k=\mbox{\boldmath $u$}_kG$. We have
\begin{equation}
r_k^{(l)h}=v_k^{(l)}+e_k^{(l)}~(1 \leq l \leq n_0) .
\end{equation}
Hence it follows that
\begin{eqnarray}
\alpha_l &=& P(e_k^{(l)}=0, r_k^{(l)h}=1)+P(e_k^{(l)}=1, r_k^{(l)h}=0) \nonumber \\
&=& P(e_k^{(l)}=0, v_k^{(l)}+e_k^{(l)}=1)+P(e_k^{(l)}=1, v_k^{(l)}+e_k^{(l)}=0) \nonumber \\
&=& P(e_k^{(l)}=0, v_k^{(l)}=1)+P(e_k^{(l)}=1, v_k^{(l)}=1) \nonumber \\
&=& P(v_k^{(l)}=1) .
\end{eqnarray}
\end{IEEEproof}
\par
Thus the distribution of $r_k^{(l)}~(1 \leq l \leq n_0)$ is given by
\begin{equation}
p_r(y)=P(v_k^{(l)}=0)q(y-c)+P(v_k^{(l)}=1)q(y+c) .
\end{equation}
This equation means that if the code symbol is $0$, then the associated distribution obeys $q(y-c)$, whereas if the code symbol is $1$, then the associated distribution obeys $q(y+c)$. Hence the result is quite reasonable. On the other hand, since the distribution of $r_k^{(l)}$ is given by the above equation, $r_k^{(l)}~(1 \leq l \leq n_0)$ are not mutually independent, because $v_k^{(l)}~(1 \leq l \leq n_0)$ are not mutually independent.
\par
Next, consider a QLI code whose generator matrix is given by
\begin{equation}
G(D)=(g_1(D), g_1(D)+D^L)~(1 \leq L \leq \nu-1) .
\end{equation}
Let $\mbox{\boldmath $\eta$}_{k-L}=(\eta_{k-L}^{(1)}, \eta_{k-L}^{(2)})$ be the input to the main decoder in an SST Viterbi decoder~\cite{taji 19}. We see that almost the same argument applies to $\beta~(=\beta_l)$ in~\cite[Proposition 14]{taji 19}. We have the following.
\begin{lem}[\cite{taji 20}]
\begin{equation}
\beta_l=P(v_k^{(l)}=1)~(l=1, 2)
\end{equation}
holds, where $\mbox{\boldmath $v$}_k=(v_k^{(1)}, v_k^{(2)})$ is the encoded block for the main decoder.
\end{lem}
\begin{IEEEproof}
In the case of QLI codes~\cite[Section II-B]{taji 19}, the hard-decision input to the main decoder is expressed as
\begin{eqnarray}
\mbox{\boldmath $\eta$}_{k-L}^h &=& u_kG+\mbox{\boldmath $e$}_{k-L} \nonumber \\
&=& \mbox{\boldmath $v$}_k+\mbox{\boldmath $e$}_{k-L} , \nonumber
\end{eqnarray}
where $u_k=\mbox{\boldmath $e$}_kF$ ($F\stackrel{\triangle}{=}\left(
\begin{array}{c}
1 \\
1
\end{array}
\right)$) and $\mbox{\boldmath $v$}_k=u_kG$. Let $\zeta_k=\mbox{\boldmath $z$}_kH(D)$ be the syndrome. Since $\mbox{\boldmath $\eta$}_{k-L}^h=(\zeta_k, \zeta_k)$, the above is equivalent to
\begin{equation}
(\zeta_k, \zeta_k)=(v_k^{(1)}, v_k^{(2)})+(e_{k-L}^{(1)}, e_{k-L}^{(2)}) .
\end{equation}
Hence we have
\begin{eqnarray}
\beta_l &=& P(e_{k-L}^{(l)}=0, \zeta_k=1)+P(e_{k-L}^{(l)}=1, \zeta_k=0) \nonumber \\
&=& P(e_{k-L}^{(l)}=0, v_k^{(l)}+e_{k-L}^{(l)}=1)+P(e_{k-L}^{(l)}=1, v_k^{(l)}+e_{k-L}^{(l)}=0) \nonumber \\
&=& P(e_{k-L}^{(l)}=0, v_k^{(l)}=1)+P(e_{k-L}^{(l)}=1, v_k^{(l)}=1) \nonumber \\
&=& P(v_k^{(l)}=1)
\end{eqnarray}
for $l=1,~2$.
\end{IEEEproof}
\par
In the rest of the paper, $n_0=2$ is assumed, because we are concerned with QLI codes in principle. Let us examine the relationship between $\alpha_l$ and $\beta_l$. Let $\hat i(k-L \vert k-L)$ and $\hat i(k-L \vert k)$ be the filtered estimate and the smoothed estimate, respectively~\cite[Section II-B]{taji 19}. We have
\begin{eqnarray}
\mbox{\boldmath $r$}_{k-L}^h &=& \mbox{\boldmath $z$}_{k-L}^h-\hat i(k-L \vert k-L)G(D) \nonumber \\
&=& (i_{k-L}-\hat i(k-L \vert k-L))G(D)+\mbox{\boldmath $e$}_{k-L} \nonumber \\
&=& u_{k-L}G(D)+\mbox{\boldmath $e$}_{k-L} \\
\mbox{\boldmath $\eta$}_{k-L}^h &=& \mbox{\boldmath $z$}_{k-L}^h-\hat i(k-L \vert k)G(D) \nonumber \\
&=& (i_{k-L}-\hat i(k-L \vert k))G(D)+\mbox{\boldmath $e$}_{k-L} \nonumber \\
&=& \tilde u_{k-L}G(D)+\mbox{\boldmath $e$}_{k-L} ,
\end{eqnarray}
where $\tilde u_{k-L}\stackrel{\triangle}{=}\mbox{\boldmath $e$}_kF$. Then it follows that
\begin{eqnarray}
\mbox{\boldmath $v$}_{k-L} &=& u_{k-L}G(D) \nonumber \\
&=& (i_{k-L}-\hat i(k-L \vert k-L))G(D) \\
\mbox{\boldmath $\tilde v$}_{k-L} &=& \tilde u_{k-L}G(D) \nonumber \\
&=& (i_{k-L}-\hat i(k-L \vert k))G(D) .
\end{eqnarray}
From the meaning of filtering and smoothing, it is natural to think that $P(i_{k-L}-\hat i(k-L \vert k-L)=0)$ is smaller than $P(i_{k-L}-\hat i(k-L \vert k)=0)$. Hence it is expected that
\begin{equation}
P(v_{k-L}^{(l)}=1) > P(\tilde v_{k-L}^{(l)}=1) ,
\end{equation}
i.e., $\alpha_l > \beta_l$ holds. (In the derivation, $P(v_{k-L}^{(l)}=1)=P(v_k^{(l)}=1)$, which is equivalent to a kind of stationarity, has been used.) However, this is not always true (see Section IV).
\par
Since the meaning of $\alpha~(=\alpha_l)$ has been clarified, we can derive the joint distribution (denoted $p_r(x, y)$) of $r_k^{(1)}$ and $r_k^{(2)}$. In fact, we have the following.
\newtheorem{pro}{Proposition}
\begin{pro}[\cite{taji 20}]
$p_r(x, y)$ is given by
\begin{eqnarray}
p_r(x, y) &=& \alpha_{00}q(x-c)q(y-c)+\alpha_{01}q(x-c)q(y+c) \nonumber \\
&& +\alpha_{10}q(x+c)q(y-c)+\alpha_{11}q(x+c)q(y+c) ,
\end{eqnarray}
where $\alpha_{ij}=P(v_k^{(1)}=i, v_k^{(2)}=j)$.
\end{pro}
\begin{IEEEproof}
$p_r(x, y) \geq 0$ is obvious. Let us show that $\int_{-\infty}^{\infty}\int_{-\infty}^{\infty}p_r(x, y)dxdy=1$. Noting, for example,
\begin{eqnarray}
\lefteqn{\int_{-\infty}^{\infty}\int_{-\infty}^{\infty}q(x-c)q(y-c)dxdy} \nonumber \\
&& =\int_{-\infty}^{\infty}q(x-c)dx\int_{-\infty}^{\infty}q(y-c)dy \nonumber \\
&& =1 \times 1=1 , \nonumber
\end{eqnarray}
we have
\begin{eqnarray}
\lefteqn{\int_{-\infty}^{\infty}\int_{-\infty}^{\infty}p_r(x, y)dxdy} \nonumber \\
&& =\alpha_{00}+\alpha_{01}+\alpha_{10}+\alpha_{11}=1 .
\end{eqnarray}
\par
({\it Remark:} $\int_{-\infty}^{\infty}\int_{-\infty}^{\infty}q(x-c)q(y-c)dxdy$ is a multiple integral on the infinite interval, i.e., an improper integral. Consider a finite interval $I=\{a \leq x \leq b, a' \leq y \leq b'\}$. Since $q(x-c)q(y-c)$ is continuous on $I$, a repeated integral is possible on $I$ and we have
\begin{eqnarray}
\lefteqn{\int\int_Iq(x-c)q(y-c)dxdy} \nonumber \\
&& =\int_a^bq(x-c)dx\int_{a'}^{b'}q(y-c)dy . \nonumber
\end{eqnarray}
Taking the limit as $a, a' \rightarrow -\infty$ and as $b, b' \rightarrow \infty$, the above equality is obtained.)
\par
Next, let us calculate the marginal distribution of $p_r(x, y)$. We have
\begin{eqnarray}
\lefteqn{\int_{-\infty}^{\infty}p_r(x, y)dy} \nonumber \\
&& =(\alpha_{00}+\alpha_{01})q(x-c)+(\alpha_{10}+\alpha_{11})q(x+c) \nonumber \\
&& =(1-\alpha_1)q(x-c)+\alpha_1q(x+c) .
\end{eqnarray}
Note that this is just the distribution of $r_k^{(1)}$. Similarly, we have
\begin{equation}
\int_{-\infty}^{\infty}p_r(x, y)dx=(1-\alpha_2)q(y-c)+\alpha_2q(y+c) ,
\end{equation}
where the right-hand side is the distribution of $r_k^{(2)}$. All these facts show that $p_r(x, y)$ is the joint distribution of $r_k^{(1)}$ and $r_k^{(2)}$.
\end{IEEEproof}

% end of file

\section{Mean-Square Error in Estimating the Input}
Consider the signal model:
\begin{equation}
Z=X+W ,
\end{equation}
where $X$ represents a signal of interest and $W$ represents random noise. We assume that $X$ and $W$ are mutually independent. Since we cannot know the value of $X$ directly, we have to estimate it based on the observation $Z$. The error of an estimate, denoted $\mbox{\boldmath $f$}(Z)$, of the input $X$ based on the observation $Z$ can be measured in mean-square sense
\begin{equation}
E[(X-\mbox{\boldmath $f$}(Z))(X-\mbox{\boldmath $f$}(Z))^T] ,
\end{equation}
where ``$^T$'' means transpose. It is well known~\cite{kai 98,kuni 76,oksen 98,will 91} that the minimum of the above value is achieved by the conditional expectation
\begin{equation}
\hat X=\widehat {\mbox{\boldmath $f$}(Z)}=E[X \vert Z] .
\end{equation}
$\hat X$ is the {\it least-squares estimate} and the corresponding estimation error is referred to as the {\it minimum mean-square error} (MMSE)~\cite{guo 05}. In the following, the value of the MMSE is denoted by ``mmse''.
\par
{\it Remark:} Let $\mathcal{Z}$ be the $\sigma$-field generated by $Z$. Also, denote by $L^2(\mathcal{Z})$ the set of elements in $L^2$ which are $\mathcal{Z}$ measurable, where $L^2$ is the set of square integrable random variables. Then we have $E[X \vert Z]=P_{L^2(\mathcal{Z})}X$, where $P_{L^2(\mathcal{Z})}X$ is the orthogonal projection of $X$ onto the space $L^2(\mathcal{Z})$~\cite{kuni 76,oksen 98,will 91}. If $X$ and $Z$ are jointly Gaussian, then we have $E[X \vert Z]=P_{\mathcal{H}(Z)}X$~\cite{kuni 76,oksen 98}, where $\mathcal{H}(Z)$ is the Gaussian space~\cite{kuni 76,oksen 98} generated by $Z$. Note that $\mathcal{H}(Z)$ is a subspace of $L^2(\mathcal{Z})$.
\par
Kailath~\cite{kai 681,kai 682} applied the innovations method to linear filtering/smoothing problems. Suppose that the observations are given by
\begin{equation}
\mbox{\boldmath $z$}_k=\mbox{\boldmath $s$}_k+\mbox{\boldmath $w$}_k,~k=1, 2, \cdots ,
\end{equation}
where $\{\mbox{\boldmath $s$}_k\}$ is a zero-mean finite-variance signal process and $\{\mbox{\boldmath $w$}_k\}$ is a zero-mean white Gaussian noise. It is assumed that $\mbox{\boldmath $w$}_k$ has a covariance matrix $E[\mbox{\boldmath $w$}_k^T\mbox{\boldmath $w$}_l]=R_k\delta_{kl}$, where $\delta_{kl}$ is Kronecker's delta. The innovation process is defined by
\begin{equation}
\mbox{\boldmath $\nu$}_k\stackrel{\triangle}{=}\mbox{\boldmath $z$}_k-\mbox{\boldmath $\hat s$}(k \vert k-1) ,
\end{equation}
where $\mbox{\boldmath $\hat s$}(k \vert k-1)$ is the linear least-squares estimate of $\mbox{\boldmath $s$}_k$ given $\{\mbox{\boldmath $z$}_l,~1 \leq l \leq k-1\}$. Kailath~\cite[Section III-B]{kai 681} showed the following.
\begin{pro}[Kailath~\cite{kai 681}]
The covariance matrix of $\mbox{\boldmath $\nu$}_k$ is given by
\begin{equation}
E[\mbox{\boldmath $\nu$}_k^T\mbox{\boldmath $\nu$}_l]=(P_k+R_k)\delta_{kl} ,
\end{equation}
where $P_k$ is the covariance matrix of the error in the estimate $\mbox{\boldmath $\hat s$}(k \vert k-1)$.
\end{pro}
\par
We remark that this result plays an essential role in this paper.

\subsection{Covariance Matrix Associated with the Input to the Main Decoder}
In this section, by assuming that $\mbox{\boldmath $r$}_k$ is the innovation, we calculate the covariance matrix of $\mbox{\boldmath $r$}_k$. Then the MMSE in estimating the input $\mbox{\boldmath $x$}_k$ is obtained from the associated covariance matrix. In preparation for the purpose, we present a lemma.
\par
\begin{lem}[\cite{taji 20}]
Suppose that $0 \leq \epsilon \leq 1/2$. The following quantities have the same value:
\begin{eqnarray}
P(v_k^{(1)}=0, v_k^{(2)}=0)-P(v_k^{(1)}=0)P(v_k^{(2)}=0) &=& \alpha_{00}-(1-\alpha_1)(1-\alpha_2) \\
P(v_k^{(1)}=0)P(v_k^{(2)}=1)-P(v_k^{(1)}=0, v_k^{(2)}=1) &=& (1-\alpha_1)\alpha_2-\alpha_{01} \\
P(v_k^{(1)}=1)P(v_k^{(2)}=0)-P(v_k^{(1)}=1, v_k^{(2)}=0) &=& \alpha_1(1-\alpha_2)-\alpha_{10} \\
P(v_k^{(1)}=1, v_k^{(2)}=1)-P(v_k^{(1)}=1)P(v_k^{(2)}=1) &=& \alpha_{11}-\alpha_1 \alpha_2 .
\end{eqnarray}
The common value is denoted by $\delta$.
\end{lem}
\par
{\it Remark:} $\delta=0$ implies that $v_k^{(1)}$ and $v_k^{(2)}$ are mutually independent.
\begin{IEEEproof}
Suppose that $\alpha_1$ and $\alpha_2$ are given. From the definition of $\alpha_{ij}$, we obtain a system of linear equations:
\begin{equation}
\left\{
\begin{array}{l}
\alpha_{00}+\alpha_{01}=1-\alpha_1 \\
\alpha_{10}+\alpha_{11}=\alpha_1 \\
\alpha_{00}+\alpha_{10}=1-\alpha_2 \\
\alpha_{01}+\alpha_{11}=\alpha_2 .
\end{array} \right.
\end{equation}
This can be solved as
\begin{equation}
\left\{
\begin{array}{l}
\alpha_{00}=1-\alpha_1-\alpha_2+u \\
\alpha_{01}=\alpha_2-u \\
\alpha_{10}=\alpha_1-u \\
\alpha_{11}=u ,
\end{array} \right.
\end{equation}
where $u$ is an arbitrary constant. We remark that the probabilities $(0 \leq )~\alpha_{00} \sim \alpha_{11}~(\leq 1)$ are determined by $u$, which in turn restricts the value of $u$. Since $0 \leq \alpha_{ij}$, $u$ must satisfy the following:
\begin{equation}
\left\{
\begin{array}{l}
u \geq \alpha_1+\alpha_2-1 \\
u \leq \alpha_2 \\
u \leq \alpha_1 \\
u \geq 0 .
\end{array} \right.
\end{equation}
Note that $0 \leq \alpha_i \leq 1/2~(i=1, 2)$ for $0 \leq \epsilon \leq 1/2$~\cite[Lemma 13]{taji 19}. Hence we have $\alpha_1+\alpha_2-1 \leq 0$. Accordingly, the value of $u$ is restricted to
\begin{equation}
0 \leq u \leq \mbox{min}(\alpha_1, \alpha_2) .
\end{equation}
It is shown that $\alpha_{ij} \leq 1$ is also satisfied for $0 \leq u \leq \mbox{min}(\alpha_1, \alpha_2)$.
\par
Now we have
\begin{eqnarray}
\lefteqn{\alpha_{00}-(1-\alpha_1)(1-\alpha_2)} \nonumber \\
&& =(1-\alpha_1-\alpha_2+u)-(1-\alpha_1-\alpha_2+\alpha_1\alpha_2) \nonumber \\
&& =u-\alpha_1\alpha_2 \nonumber \\
&& =\alpha_{11}-\alpha_1\alpha_2 .
\end{eqnarray}
We see that the remaining three quantities are equal also to $u-\alpha_1\alpha_2=\alpha_{11}-\alpha_1\alpha_2$.
\end{IEEEproof}
\par
Let us show that $\delta \geq 0$. We need the following.
\begin{lem}
\begin{equation}
P(e_1+e_2+\cdots +e_m=1) \leq \frac{1}{2}
\end{equation}
holds, where errors $e_j~(1 \leq j \leq m)$ are statistically independent of each other.
\end{lem}
\begin{IEEEproof}
See Appendix A.
\end{IEEEproof}
\par
Now we have the following.
\begin{lem}
Suppose that $0 \leq \epsilon \leq 1/2$. Then $\delta \geq 0$ holds.
\end{lem}
\begin{IEEEproof}
See Appendix B.
\end{IEEEproof}
\par
\begin{pro}[\cite{taji 20}]
The covariance matrix associated with $p_r(x, y)$ is given by
\begin{eqnarray}
\Sigma_r &\stackrel{\triangle}{=}& \left(
\begin{array}{cc}
\sigma_{r_1}^2 & \sigma_{r_1r_2} \\
\sigma_{r_1r_2} & \sigma_{r_2}^2
\end{array}
\right) \nonumber \\
&=& \left(
\begin{array}{cc}
1+4c^2\alpha_1(1-\alpha_1) & 4c^2 \delta \\
4c^2 \delta & 1+4c^2\alpha_2(1-\alpha_2)
\end{array}
\right) .
\end{eqnarray}
\end{pro}
\begin{IEEEproof}
Note the relation
\begin{equation}
\sigma_{r_1}^2=\int_{-\infty}^{\infty}\int_{-\infty}^{\infty}x^2\,p_r(x, y)dxdy-m_{r_1}^2 ,
\end{equation}
where
\begin{equation}
m_{r_1}=\int_{-\infty}^{\infty}\int_{-\infty}^{\infty}x\,p_r(x, y)dxdy .
\end{equation}
The first term is calculated as
\begin{eqnarray}
\lefteqn{\int_{-\infty}^{\infty}\int_{-\infty}^{\infty}x^2\,p_r(x, y)dxdy} \nonumber \\
&& =(\alpha_{00}+\alpha_{01}+\alpha_{10}+\alpha_{11})(1+c^2) \nonumber \\
&& =1+c^2 .
\end{eqnarray}
Also, we have
\begin{eqnarray}
m_{r_1} &=& \int_{-\infty}^{\infty}\int_{-\infty}^{\infty}x\,p_r(x, y)dxdy \nonumber \\
&=& (\alpha_{00}+\alpha_{01})c+(\alpha_{10}+\alpha_{11})(-c) \nonumber \\
&=& (1-\alpha_1)c-\alpha_1c \nonumber \\
&=& (1-2\alpha_1)c .
\end{eqnarray}
Then it follows that
\begin{eqnarray}
\sigma_{r_1}^2 &=& (1+c^2)-(1-2\alpha_1)^2c^2 \nonumber \\
&=& 1+4c^2\alpha_1(1-\alpha_1) .
\end{eqnarray}
Similarly, we have
\begin{eqnarray}
\sigma_{r_2}^2 &=& \int_{-\infty}^{\infty}\int_{-\infty}^{\infty}y^2\,p_r(x, y)dxdy-m_{r_2}^2 \nonumber \\
&=& 1+4c^2\alpha_2(1-\alpha_2) .
\end{eqnarray}
\par
Let us calculate $\sigma_{r_1r_2}$. Note this time the relation
\begin{equation}
\sigma_{r_1r_2}=\int_{-\infty}^{\infty}\int_{-\infty}^{\infty}xy\,p_r(x, y)dxdy-m_{r_1}m_{r_2} .
\end{equation}
Applying Lemma 3, we have
\begin{eqnarray}
\lefteqn{\int_{-\infty}^{\infty}\int_{-\infty}^{\infty}xy\,p_r(x, y)dxdy} \nonumber \\
&& =(\alpha_{00}+\alpha_{11})c^2-(\alpha_{01}+\alpha_{10})c^2 \nonumber \\
&& =(1-\alpha_1-\alpha_2+2u)c^2-(\alpha_1+\alpha_2-2u)c^2 \nonumber \\
&& =(1-2\alpha_1-2\alpha_2+4u)c^2 .
\end{eqnarray}
Since
\begin{equation}
\left\{
\begin{array}{l}
m_{r_1}=(1-2\alpha_1)c \\
m_{r_2}=(1-2\alpha_2)c ,
\end{array} \right.
\end{equation}
it follows that
\begin{eqnarray}
\sigma_{r_1r_2} &=& (1-2\alpha_1-2\alpha_2+4u)c^2-(1-2\alpha_1)(1-2\alpha_2)c^2 \nonumber \\
&=& 4(u-\alpha_1\alpha_2)c^2 \nonumber \\
&=& 4\delta c^2 .
\end{eqnarray}
\end{IEEEproof}

\subsection{MMSE in Estimating the Input}
Note Proposition 2. If $\mbox{\boldmath $r$}_k=(r_k^{(1)}, r_k^{(2)})$ corresponds to the innovation, then the associated covariance matrix (denoted $\Sigma_r$) is expressed as the sum of two matrices $\Sigma_x$ and $\Sigma_w$, i.e., $\Sigma_r=\Sigma_x+\Sigma_w$. Here, $\Sigma_x$ is the covariance matrix of the estimation error of the signal, whereas $\Sigma_w$ is the covariance matrix of the observation noise. From the definition of $w_j$ (see Section I), it follows that
\begin{equation}
\Sigma_w=\left(
\begin{array}{cc}
1 & 0 \\
0 & 1
\end{array}
\right)
\end{equation}
and hence we have
\begin{equation}
\Sigma_x=\left(
\begin{array}{cc}
4c^2\alpha_1(1-\alpha_1) & 4c^2 \delta \\
4c^2 \delta & 4c^2\alpha_2(1-\alpha_2)
\end{array}
\right) .
\end{equation}
Moreover, since the signal is expressed as $c\mbox{\boldmath $x$}_k$, the covariance matrix (denoted $\hat \Sigma_x$) of the estimation error of $\mbox{\boldmath $x$}_k$ becomes
\begin{equation}
\hat \Sigma_x=\left(
\begin{array}{cc}
4\alpha_1(1-\alpha_1) & 4 \delta \\
4 \delta & 4\alpha_2(1-\alpha_2)
\end{array}
\right) .
\end{equation}
Hence the corresponding MMSE is given by
\begin{equation}
\mbox{mmse}=\mbox{tr}\,\hat \Sigma_x=4\alpha_1(1-\alpha_1)+4\alpha_2(1-\alpha_2) ,
\end{equation}
where $\mbox{tr}\,\hat \Sigma_x$ is the trace of matrix $\hat \Sigma_x$.
\par
{\it Remark:} The estimate in~\cite{kai 681} is the ``linear'' least-squares estimate. On the other hand, the best estimate is given by the conditional expectation $E[\mbox{\boldmath $x$}_k \vert \mbox{\boldmath $z$}_l,~1 \leq l \leq k-1]$, which is in general a highly ``nonlinear'' functional of $\{\mbox{\boldmath $z$}_l,~1 \leq l \leq k-1\}$ (cf. $\mbox{\boldmath $x$}_k$ is not Gaussian). Note that if $\{\mbox{\boldmath $x$}_k, \mbox{\boldmath $z$}_k\}$ are jointly Gaussian, then $E[\mbox{\boldmath $x$}_k \vert \mbox{\boldmath $z$}_l,~1 \leq l \leq k-1]$ is a linear functional of $\{\mbox{\boldmath $z$}_l,~1 \leq l \leq k-1\}$~\cite[Section II-D]{kai 98}. Hence to be exact, we have
\begin{equation}
\mbox{mmse} \leq 4\alpha_1(1-\alpha_1)+4\alpha_2(1-\alpha_2) ,
\end{equation}
where the left-hand side corresponds to the conditional expectation, whereas the right-hand side corresponds to the linear least-squares estimate. As stated above, the MMSE is a nonlinear functional of the past observations. Hence we regard the right-hand side as an approximation to the MMSE.

\subsection{Modification of the Derived MMSE}
Recall that $\alpha_1=P(v_k^{(1)}=1)$ and $\alpha_2=P(v_k^{(2)}=1)$. That is, the MMSE in the estimation of $\mbox{\boldmath $x$}_k$ is expressed in terms of the distribution of $\mbox{\boldmath $v$}_k=(v_k^{(1)}, v_k^{(2)})$ for the main decoder in an SST Viterbi decoder. We remark that $\alpha_1$ and $\alpha_2$ are not mutually independent. As a result, $4\alpha_1(1-\alpha_1)$ and $4\alpha_2(1-\alpha_2)$ have a correlation. Hence the simple sum of $4\alpha_1(1-\alpha_1)$ and $4\alpha_2(1-\alpha_2)$ is not appropriate for the MMSE and some modification considering the degree of correlation is necessary. The variable $\delta$ (see Lemma 3) has a close connection with the independence of $v_k^{(1)}$ and $v_k^{(2)}$. When they have a weak correlation, $\delta$ is small, whereas when they have a strong correlation, $\delta$ is large. This observation suggests that a modification
\begin{equation}
4\alpha_1(1-\alpha_1)+4\alpha_2(1-\alpha_2)-\lambda \delta
\end{equation}
is more appropriate for the MMSE, where $\lambda$ is some constant. In the following, we discuss how to determine the value of $\lambda$. We have the following.
\begin{pro}
Suppose that $0 \leq \epsilon \leq 1/2$. We have
\begin{equation}
(4\alpha_1(1-\alpha_1))(4\alpha_2(1-\alpha_2)) \geq (4 \delta)^2 ,
\end{equation}
where $\delta=\alpha_{11}-\alpha_1\alpha_2$.
\end{pro}
\begin{IEEEproof}
Note the relation: $\Sigma_x=\Sigma_r+\Sigma_w$. $\Sigma_r$ is the covariance matrix associated with $p_r(x, y)$ and is positive semi-definite. $\Sigma_w$ (i.e., the identity matrix of size $2 \times 2$) is clearly positive semi-definite. Hence $\Sigma_x$ is positive semi-definite and hence $\mbox{det}(\Sigma_x) \geq 0$~\cite{str 76}, where ``$\mbox{det}(\cdot)$'' denotes the determinant. Since $\mbox{det}(\Sigma_x)=c^4\mbox{det}(\hat \Sigma_x)$, we have $\mbox{det}(\hat \Sigma_x) \geq 0$.
\end{IEEEproof}
\par
From Proposition 4, we have
\begin{displaymath}
(4\alpha_1(1-\alpha_1))(4\alpha_2(1-\alpha_2)) \geq (4 \delta)^2 .
\end{displaymath}
So far we have carried out numerical calculations for four QLI codes $C_1 \sim C_4$ (regarded as general codes) and one general code $C_5$ (these codes are defined in Sections IV and V). Then we have found that the difference between the values of $4\alpha_1(1-\alpha_1)$ and $4\alpha_2(1-\alpha_2)$ is small. Note that this fact is derived from the structure of $G^{-1}G$. Let
\begin{equation}
G^{-1}G=\left(
\begin{array}{cc}
b_{11} & b_{12} \\
b_{21} & b_{22}
\end{array}
\right) .
\end{equation}
Also, denote by $m_{col}^{(i)}~(i=1, 2)$ the number of terms (i.e., $D^j$) contained in $\left(
\begin{array}{c}
b_{1i} \\
b_{2i}
\end{array}
\right)$. We see that $\alpha_i~(i=1, 2)$ is determined by $m_{col}^{(i)}$. Hence if the difference between $m_{col}^{(1)}$ and $m_{col}^{(2)}$ is small, then $\alpha_1 \approx \alpha_2$ holds. For example, it is shown (see Section IV-B) that
\begin{itemize}
\item [1)] $m_{col}^{(1)}=5$ and $m_{col}^{(2)}=6$ for $C_1$,
\item [2)] $m_{col}^{(1)}=10$ and $m_{col}^{(2)}=10$ for $C_2$.
\end{itemize}
Thus we have $4\alpha_1(1-\alpha_1) \approx 4\alpha_2(1-\alpha_2)$. As a consequence, we have approximately
\begin{equation}
\left\{
\begin{array}{l}
4\alpha_1(1-\alpha_1) \geq 4 \delta \\
4\alpha_2(1-\alpha_2) \geq 4 \delta .
\end{array} \right.
\end{equation}
That is,
\begin{displaymath}
4\alpha_1(1-\alpha_1)+4\alpha_2(1-\alpha_2)-8\delta \geq 0
\end{displaymath}
holds with high probability.
\par
Now we can show that this inequality always holds. That is, we have the following.
\begin{pro}
Suppose that $0 \leq \epsilon \leq 1/2$. We have
\begin{equation}
4\alpha_1(1-\alpha_1)+4\alpha_2(1-\alpha_2)-8\delta \geq 0 ,
\end{equation}
where $\delta=\alpha_{11}-\alpha_1\alpha_2$.
\end{pro}
\begin{IEEEproof}
See Appendix C.
\end{IEEEproof}
\par
{\it Remark:} Let
\begin{eqnarray}
\hat \Sigma_x &=& \left(
\begin{array}{cc}
4\alpha_1(1-\alpha_1) & 4 \delta \\
4 \delta & 4\alpha_2(1-\alpha_2)
\end{array}
\right) \nonumber \\
&\stackrel{\triangle}{=}& \left(
\begin{array}{cc}
\sigma_1^2 & \sigma_{12} \\
\sigma_{12} & \sigma_2^2
\end{array}
\right) .
\end{eqnarray}
Then $4\alpha_1(1-\alpha_1)+4\alpha_2(1-\alpha_2)-8\delta$ corresponds to $\sigma_1^2+\sigma_2^2-2\sigma_{12}$. Denote by $\mu$ the correlation coefficient between $x_k^{(1)}$ and $x_k^{(2)}$, i.e.,
\begin{equation}
\mu\stackrel{\triangle}{=}\frac{\sigma_{12}}{\sigma_1\sigma_2} .
\end{equation}
Note that $-1 \leq \mu \leq 1$. We have
\begin{equation}
\sigma_1^2+\sigma_2^2-2\sigma_{12}=\sigma_1^2+\sigma_2^2-2\mu\sigma_1\sigma_2 .
\end{equation}
Now we restrict $\mu$ to $0 \leq \mu \leq 1$. As the special cases, the following hold:
\begin{itemize}
\item [1)] $\mu=0$: $\sigma_1^2+\sigma_2^2-2\mu\sigma_1\sigma_2=\sigma_1^2+\sigma_2^2$.
\item [2)] $\mu=1$: $\sigma_1^2+\sigma_2^2-2\mu\sigma_1\sigma_2=(\sigma_1-\sigma_2)^2$.
\end{itemize}
Hence $-8\delta=-2\mu\sigma_1\sigma_2$ represents the correction term depending on the degree of correlation.
\par
Based on Proposition 5, we finally set
\begin{eqnarray}
\mbox{mmse} &=& 4\alpha_1(1-\alpha_1)+4\alpha_2(1-\alpha_2)-8\delta \\
&\stackrel{\triangle}{=}& \xi(\alpha_1, \alpha_2, \delta)
\end{eqnarray}
as the MMSE in the estimation of $\mbox{\boldmath $x$}_k$.

\subsection{In the Case of QLI Codes}
Consider a QLI code whose generator matrix is given by
\begin{displaymath}
G(D)=(g_1(D), g_1(D)+D^L)~(1 \leq L \leq \nu-1) .
\end{displaymath}
Let $\mbox{\boldmath $\eta$}_{k-L}=(\eta_{k-L}^{(1)}, \eta_{k-L}^{(2)})$ be the input to the main decoder in an SST Viterbi decoder. We see that the results in the previous sections hold for $\mbox{\boldmath $\eta$}_{k-L}$ as well. Therefore, we state only the results.
\par
\begin{pro}[\cite{taji 20}]
The joint distribution of $\eta_{k-L}^{(1)}$ and $\eta_{k-L}^{(2)}$ (denoted $p_{\eta}(x, y)$) is given by
\begin{eqnarray}
p_{\eta}(x, y) &=& \beta_{00}q(x-c)q(y-c)+\beta_{01}q(x-c)q(y+c) \nonumber \\
&& +\beta_{10}q(x+c)q(y-c)+\beta_{11}q(x+c)q(y+c) ,
\end{eqnarray}
where $\beta_{ij}=P(v_k^{(1)}=i, v_k^{(2)}=j)$.
\end{pro}
\begin{lem}[\cite{taji 20}]
Assume that $0 \leq \epsilon \leq 1/2$. The following quantities have the same value:
\begin{eqnarray}
P(v_k^{(1)}=0, v_k^{(2)}=0)-P(v_k^{(1)}=0)P(v_k^{(2)}=0) &=& \beta_{00}-(1-\beta_1)(1-\beta_2) \\
P(v_k^{(1)}=0)P(v_k^{(2)}=1)-P(v_k^{(1)}=0, v_k^{(2)}=1) &=& (1-\beta_1)\beta_2-\beta_{01} \\
P(v_k^{(1)}=1)P(v_k^{(2)}=0)-P(v_k^{(1)}=1, v_k^{(2)}=0) &=& \beta_1(1-\beta_2)-\beta_{10} \\
P(v_k^{(1)}=1, v_k^{(2)}=1)-P(v_k^{(1)}=1)P(v_k^{(2)}=1) &=& \beta_{11}-\beta_1 \beta_2 .
\end{eqnarray}
The common value is denoted by $\delta'$.
\end{lem}
\begin{pro}[\cite{taji 20}]
The covariance matrix associated with $p_{\eta}(x, y)$ is given by
\begin{eqnarray}
\Sigma_{\eta} &\stackrel{\triangle}{=}& \left(
\begin{array}{cc}
\sigma_{\eta_1}^2 & \sigma_{\eta_1\eta_2} \\
\sigma_{\eta_1\eta_2} & \sigma_{\eta_2}^2
\end{array}
\right) \nonumber \\
&=& \left(
\begin{array}{cc}
1+4c^2\beta_1(1-\beta_1) & 4c^2 \delta' \\
4c^2 \delta' & 1+4c^2\beta_2(1-\beta_2)
\end{array}
\right) .
\end{eqnarray}
\end{pro}
\begin{pro}
Suppose that $0 \leq \epsilon \leq 1/2$. We have
\begin{equation}
(4\beta_1(1-\beta_1))(4\beta_2(1-\beta_2)) \geq (4\delta')^2 ,
\end{equation}
where $\delta'=\beta_{11}-\beta_1\beta_2$.
\end{pro}
\begin{pro}
Suppose that $0 \leq \epsilon \leq 1/2$. We have
\begin{equation}
4\beta_1(1-\beta_1)+4\beta_2(1-\beta_2)-8\delta' \geq 0 ,
\end{equation}
where $\delta'=\beta_{11}-\beta_1\beta_2$.
\end{pro}
\par
Finally, for QLI codes we set
\begin{eqnarray}
\mbox{mmse} &=& 4\beta_1(1-\beta_1)+4\beta_2(1-\beta_2)-8\delta' \\
&\stackrel{\triangle}{=}& \xi(\beta_1, \beta_2, \delta')
\end{eqnarray}
as the MMSE in the estimation of $\mbox{\boldmath $x$}_{k-L}$.

% end of file

\section{Mutual Information and MMSEs}
We have shown that the MMSE in the estimation of $\mbox{\boldmath $x$}_k$ is expressed in terms of the distribution of $\mbox{\boldmath $v$}_k=(v_k^{(1)}, v_k^{(2)})$ for the main decoder in an SST Viterbi decoder. More precisely, we have derived the following:
\begin{itemize}
\item [1)] As a general code: $\mbox{mmse}=\xi(\alpha_1, \alpha_2, \delta)$.
\item [2)] As a QLI code: $\mbox{mmse}=\xi(\beta_1, \beta_2, \delta')$.
\end{itemize}
In this section, we discuss the validity of the derived MMSE from the viewpoint of mutual information and mean-square error.
\par
In this paper, we have used the signal model (see Section I):
\begin{displaymath}
\mbox{\boldmath $z$}_k=c \mbox{\boldmath $x$}_k+\mbox{\boldmath $w$}_k .
\end{displaymath}
Since $n_0=2$, it follows that
\begin{equation}
c=\sqrt{2E_s/N_0}=\sqrt{E_b/N_0} .
\end{equation}
Then letting
\begin{equation}
\rho=c^2=E_b/N_0 ,
\end{equation}
the above equation is rewritten as
\begin{equation}
\mbox{\boldmath $z$}_k=\sqrt{\rho}\,\mbox{\boldmath $x$}_k+\mbox{\boldmath $w$}_k .
\end{equation}
Note that this is just the signal model used in~\cite{guo 05} (i.e., $\rho=\mbox{snr}$). Guo et al.~\cite{guo 05} discussed the relation between the input-output mutual information and the MMSE in estimating the input in Gaussian channels. Their relation holds for discrete-time and continuous-time noncausal (smoothing) MMSE estimation regardless of the input statistics (i.e., not necessarily Gaussian). Here~\cite{taji 19} recall that the innovations approach to Viterbi decoding of QLI codes has a close connection with smoothing in the linear estimation theory. Then we thought the results in~\cite{guo 05} can be used to discuss the validity of the MMSE obtained in the previous section. Hence the argument in this section is based on that of Guo et al.~\cite{guo 05}.

\subsection{General Codes}
Let $\mbox{\boldmath $z$}^n\stackrel{\triangle}{=}\{\mbox{\boldmath $z$}_1, \mbox{\boldmath $z$}_2, \cdots, \mbox{\boldmath $z$}_n\}$. Also, let $E[\mbox{\boldmath $x$} \vert \mbox{\boldmath $z$}^n]$ be the conditional expectation of {\boldmath $x$} given $\mbox{\boldmath $z$}^n$. According to Guo et al.~\cite[Section IV-A]{guo 05}, let us define as follows:
\begin{equation}
\mbox{pmmse}(i, \rho)\stackrel{\triangle}{=}E[(\mbox{\boldmath $x$}_i-E[\mbox{\boldmath $x$}_i \vert \mbox{\boldmath $z$}^{i-1}])(\mbox{\boldmath $x$}_i-E[\mbox{\boldmath $x$}_i \vert \mbox{\boldmath $z$}^{i-1}])^T] ,
\end{equation}
where $\mbox{pmmse}(i, \rho)$ represents the one-step prediction MMSE. Note that this is a function of $\rho~(=c^2)$. This is because $E[\mbox{\boldmath $x$}_i \vert \mbox{\boldmath $z$}^{i-1}]$ depends on $\mbox{\boldmath $z$}^{i-1}$ (cf. $\mbox{\boldmath $z$}^{i-1}$ is a function of $\rho$).
\par
{\it Remark 1:} The above definition is slightly different from that in~\cite{guo 05}, where $\mbox{pmmse}(i, \rho)$ is defined as
\begin{equation}
\mbox{pmmse}(i, \rho)\stackrel{\triangle}{=}E[\vert x_i-E[x_i \vert z^{i-1}]\vert^2] ,
\end{equation}
where $z^{i-1}\stackrel{\triangle}{=}\{z_1, \cdots, z_{i-1}\}$. In order to distinguish it from ours, theirs is denoted by $\mbox{pmmse}_i(\rho)$ in this section. Set $n=2k$. We have
\begin{eqnarray}
\lefteqn{\sum_{i=1}^n\mbox{pmmse}_i(\rho)} \nonumber \\
&& =\sum_{i=1}^k(\mbox{pmmse}_{2i-1}(\rho)+\mbox{pmmse}_{2i}(\rho)) .
\end{eqnarray}
Here note the following:
\begin{eqnarray}
\mbox{pmmse}_{2i-1}(\rho) &=& E[\vert x_{2i-1}-E[x_{2i-1} \vert z^{2i-2}]\vert^2] \\
\mbox{pmmse}_{2i}(\rho) &=& E[\vert x_{2i}-E[x_{2i} \vert z^{2i-1}]\vert^2] \nonumber \\
&\leq& E[\vert x_{2i}-E[x_{2i} \vert z^{2i-2}]\vert^2] .
\end{eqnarray}
Hence
\begin{eqnarray}
\mbox{pmmse}_{2i-1}(\rho)+\mbox{pmmse}_{2i}(\rho) &\leq& E[\vert x_{2i-1}-E[x_{2i-1} \vert z^{2i-2}]\vert^2] \nonumber \\
&& +E[\vert x_{2i}-E[x_{2i} \vert z^{2i-2}]\vert^2] \nonumber \\
&=& E[(\mbox{\boldmath $x$}_i-E[\mbox{\boldmath $x$}_i \vert \mbox{\boldmath $z$}^{i-1}])(\mbox{\boldmath $x$}_i-E[\mbox{\boldmath $x$}_i \vert \mbox{\boldmath $z$}^{i-1}])^T]
\end{eqnarray}
holds. Thus we have
\begin{eqnarray}
\lefteqn{\sum_{i=1}^n\mbox{pmmse}_i(\rho)} \nonumber \\
&& =\sum_{i=1}^k(\mbox{pmmse}_{2i-1}(\rho)+\mbox{pmmse}_{2i}(\rho)) \nonumber \\
&& \leq \sum_{i=1}^kE[(\mbox{\boldmath $x$}_i-E[\mbox{\boldmath $x$}_i \vert \mbox{\boldmath $z$}^{i-1}])(\mbox{\boldmath $x$}_i-E[\mbox{\boldmath $x$}_i \vert \mbox{\boldmath $z$}^{i-1}])^T] \nonumber \\
&& =\sum_{i=1}^k\mbox{pmmse}(i, \rho) .
\end{eqnarray}
\par
It is confirmed from Remark 1 that the relation in~\cite[Theorem 9]{guo 05} still holds. In fact, we have
\begin{eqnarray}
I[\mbox{\boldmath $x$}^k; \mbox{\boldmath $z$}^k] &\leq& \frac{\rho}{2}\sum_{i=1}^{2k}\mbox{pmmse}_i(\rho) \nonumber \\
&\leq& \frac{\rho}{2}\sum_{i=1}^k\mbox{pmmse}(i, \rho) .
\end{eqnarray}
Then it follows that
\begin{equation}
\frac{1}{\rho}\left(\frac{I[\mbox{\boldmath $x$}^k; \mbox{\boldmath $z$}^k]}{k}\right) \leq \frac{1}{2}\left(\frac{1}{k}\sum_{i=1}^k\mbox{pmmse}(i, \rho)\right) .
\end{equation}
\par
Our argument is based on the innovations approach proposed by Kailath~\cite{kai 681} (cf.~\cite{taji 19}). Also, our model is a discrete-time one. Hence it is reasonable to think that the MMSE associated with the estimation of $\mbox{\boldmath $x$}_i$ is corresponding to $\mbox{pmmse}(i, \rho)$ (see Proposition 2). That is, it is appropriate to set
\begin{equation}
\mbox{pmmse}(i, \rho)=\xi(\alpha_1, \alpha_2, \delta) .
\end{equation}
Moreover, we set
\begin{equation}
\frac{1}{k}\sum_{i=1}^k\mbox{pmmse}(i, \rho) \approx \xi(\alpha_1, \alpha_2, \delta) .
\end{equation}
Note that the left-hand side is the average prediction MMSE (see~\cite[Section IV-A]{guo 05}). We finally have
\begin{equation}
\frac{1}{\rho}\left(\frac{I[\mbox{\boldmath $x$}^k; \mbox{\boldmath $z$}^k]}{k}\right) \lesssim \frac{1}{2}\xi(\alpha_1, \alpha_2, \delta) .
\end{equation}
\par
On the other hand, the mutual information between $\mbox{\boldmath $x$}^k$ and $\mbox{\boldmath $z$}^k$~\cite[Section 9.4]{cov 06} is evaluated as
\begin{equation}
I[\mbox{\boldmath $x$}^k; \mbox{\boldmath $z$}^k] \leq \frac{1}{2}\log(1+\rho)^{2k}=k\log(1+\rho) .
\end{equation}
Then we have
\begin{equation}
\frac{1}{\rho}\left(\frac{I[\mbox{\boldmath $x$}^k; \mbox{\boldmath $z$}^k]}{k}\right) \leq \frac{\log(1+\rho)}{\rho} .
\end{equation}
\par
{\it Remark 2:} The essence of our argument lies in the equality
\begin{displaymath}
\mbox{pmmse}(i, \rho)=\xi(\alpha_1, \alpha_2, \delta) .
\end{displaymath}
Meanwhile, using the signal model, $\mbox{pmmse}(i, \rho)$ is calculated as follows. Since $\mbox{\boldmath $x$}_i$ and $\mbox{\boldmath $z$}^{i-1}$ are mutually independent, we have
\begin{eqnarray}
\mbox{pmmse}(i, \rho) &=& E[(\mbox{\boldmath $x$}_i-E[\mbox{\boldmath $x$}_i \vert \mbox{\boldmath $z$}^{i-1}])(\mbox{\boldmath $x$}_i-E[\mbox{\boldmath $x$}_i \vert \mbox{\boldmath $z$}^{i-1}])^T] \nonumber \\
&=& E[(\mbox{\boldmath $x$}_i-E[\mbox{\boldmath $x$}_i])(\mbox{\boldmath $x$}_i-E[\mbox{\boldmath $x$}_i])^T] \nonumber \\
&=& \mbox{var}(x_i^{(1)})+\mbox{var}(x_i^{(2)}) \nonumber \\
&=& 1+1=2 ,
\end{eqnarray}
where ``var'' denotes the variance. Hence
\begin{equation}
\frac{\rho}{2}\sum_{i=1}^k\mbox{pmmse}(i, \rho)=k \rho .
\end{equation}
On the other hand, using the inequality: $\log\,x \leq x-1~(x>0)$, we have
\begin{equation}
\frac{1}{\rho}\left(\frac{I[\mbox{\boldmath $x$}^k; \mbox{\boldmath $z$}^k]}{k}\right) \leq \frac{\log(1+\rho)}{\rho} \leq \frac{\rho}{\rho}=1 .
\end{equation}
Hence
\begin{displaymath}
I[\mbox{\boldmath $x$}^k; \mbox{\boldmath $z$}^k] \leq k\rho=\frac{\rho}{2}\sum_{i=1}^k\mbox{pmmse}(i, \rho)
\end{displaymath}
actually holds.
\par
From the above argument, it is expected that $\frac{\log(1+\rho)}{\rho}$ and $\frac{1}{2}\xi(\alpha_1, \alpha_2, \delta)$ are closely related. Then in the next section, we will discuss the relation between them.

\subsection{Numerical Results as General Codes}
In order to examine the relation between $\frac{\log(1+\rho)}{\rho}$ and $\frac{1}{2}\xi(\alpha_1, \alpha_2, \delta)$, numerical calculations have been carried out. We have taken two QLI codes:
\begin{itemize}
\item [1)] $C_1~(\nu=2, L=1)$: The generator matrix is defined by $G_1(D)=(1+D+D^2, 1+D^2)$.
\item [2)] $C_2~(\nu=6, L=2)$: The generator matrix is defined by $G_2(D)=(1+D+D^3+D^4+D^6, 1+D+D^2+D^3+D^4+D^6)$.
\end{itemize}
By regarding these QLI codes as ``general'' codes, we have compared the values of $\frac{\log(1+\rho)}{\rho}$ and $\frac{1}{2}\xi(\alpha_1, \alpha_2, \delta)$. The results are shown in Tables I and II, where $\frac{1}{2}\xi(\alpha_1, \alpha_2, \delta)$ is simply denoted $\frac{1}{2}\mbox{mmse}$. The results for $C_1$ and $C_2$ are shown also in Fig.1 and Fig.2, respectively.
\begin{table}[tb]
\caption{$\frac{\log(1+\rho)}{\rho}$ and minimum mean-square error ($C_1$ as a general code)}
\label{Table 1}
\begin{center}
\begin{tabular}{c*{7}{|c}}
$E_b/N_0~(\mbox{dB})$ & $\frac{\log(1+\rho)}{\rho}$ & $\alpha_1$ & $4\alpha_1(1-\alpha_1)$ & $\alpha_2$ & $4\alpha_2(1-\alpha_2)$ & $\delta$ & $\frac{1}{2}\mbox{mmse}$ \\
\hline
$-10$ & $0.9531$ & $0.4995$ & $1.0000$ & $0.4999$ & $1.0000$ & $0.0038$ & $0.9848$ \\
$-9$ & $0.9419$ & $0.4992$ & $1.0000$ & $0.4998$ & $1.0000$ & $0.0053$ & $0.9788$ \\
$-8$ & $0.9282$ & $0.4986$ & $1.0000$ & $0.4996$ & $1.0000$ & $0.0074$ & $0.9704$ \\
$-7$ & $0.9118$ & $0.4976$ & $1.0000$ & $0.4992$ & $1.0000$ & $0.0102$ & $0.9592$ \\
$-6$ & $0.8921$ & $0.4958$ & $0.9999$ & $0.4984$ & $1.0000$ & $0.0142$ & $0.9432$ \\
$-5$ & $0.8689$ & $0.4930$ & $0.9998$ & $0.4970$ & $1.0000$ & $0.0193$ & $0.9227$ \\
$-4$ & $0.8418$ & $0.4883$ & $0.9995$ & $0.4945$ & $0.9999$ & $0.0262$ & $0.8949$ \\
$-3$ & $0.8106$ & $0.4808$ & $0.9985$ & $0.4900$ & $0.9996$ & $0.0352$ & $0.8583$ \\
$-2$ & $0.7753$ & $0.4691$ & $0.9962$ & $0.4823$ & $0.9987$ & $0.0465$ & $0.8115$ \\
$-1$ & $0.7360$ & $0.4515$ & $0.9906$ & $0.4696$ & $0.9963$ & $0.0602$ & $0.7527$ \\
$0$ & $0.6931$ & $0.4259$ & $0.9780$ & $0.4494$ & $0.9898$ & $0.0758$ & $0.6807$ \\
$1$ & $0.6473$ & $0.3904$ & $0.9520$ & $0.4191$ & $0.9738$ & $0.0917$ & $0.5961$ \\
$2$ & $0.5992$ & $0.3442$ & $0.9029$ & $0.3766$ & $0.9391$ & $0.1050$ & $0.5010$ \\
$3$ & $0.5498$ & $0.2879$ & $0.8201$ & $0.3213$ & $0.8723$ & $0.1116$ & $0.3998$ \\
$4$ & $0.5001$ & $0.2255$ & $0.6986$ & $0.2565$ & $0.7628$ & $0.1076$ & $0.3003$ \\
$5$ & $0.4510$ & $0.1621$ & $0.5433$ & $0.1876$ & $0.6096$ & $0.0921$ & $0.2081$ \\
$6$ & $0.4033$ & $0.1049$ & $0.3756$ & $0.1231$ & $0.4318$ & $0.0681$ & $0.1313$ \\
$7$ & $0.3579$ & $0.0599$ & $0.2252$ & $0.0710$ & $0.2638$ & $0.0427$ & $0.0737$ \\
$8$ & $0.3153$ & $0.0239$ & $0.1138$ & $0.0349$ & $0.1347$ & $0.0222$ & $0.0355$ \\
$9$ & $0.2758$ & $0.0120$ & $0.0474$ & $0.0143$ & $0.0564$ & $0.0094$ & $0.0143$ \\
$10$ & $0.2398$ & $0.0039$ & $0.0155$ & $0.0047$ & $0.0187$ & $0.0031$ & $0.0047$
\end{tabular}
\end{center}
\end{table}
\begin{figure}[tb]
\begin{center}
\includegraphics[width=10.0cm,clip]{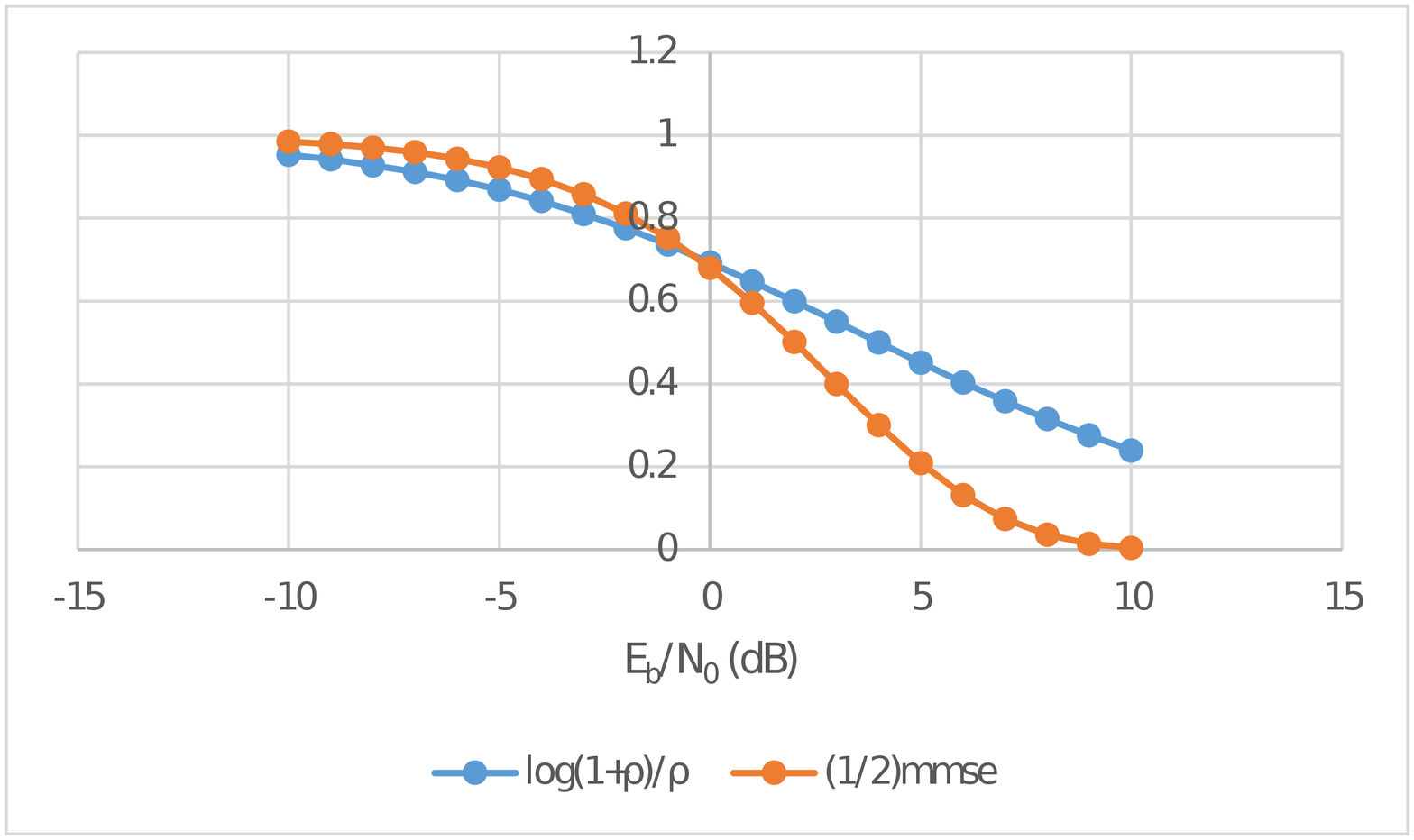}
\end{center}
\caption{$\frac{\log(1+\rho)}{\rho}$ and $\frac{1}{2}\mbox{mmse}=\frac{1}{2}\xi(\alpha_1, \alpha_2, \delta)~(C_1)$.}
\label{Fig.1}
\end{figure}
\begin{table}[tb]
\caption{$\frac{\log(1+\rho)}{\rho}$ and minimum mean-square error ($C_2$ as a general code)}
\label{Table 2}
\begin{center}
\begin{tabular}{c*{7}{|c}}
$E_b/N_0~(\mbox{dB})$ & $\frac{\log(1+\rho)}{\rho}$ & $\alpha_1$ & $4\alpha_1(1-\alpha_1)$ & $\alpha_2$ & $4\alpha_2(1-\alpha_2)$ & $\delta$ & $\frac{1}{2}\mbox{mmse}$ \\
\hline
$-10$ & $0.9531$ & $0.5000$ & $1.0000$ & $0.5000$ & $1.0000$ & $0.0000$ & $1.0000$ \\
$-9$ & $0.9419$ & $0.5000$ & $1.0000$ & $0.5000$ & $1.0000$ & $0.0000$ & $1.0000$ \\
$-8$ & $0.9282$ & $0.5000$ & $1.0000$ & $0.5000$ & $1.0000$ & $0.0000$ & $1.0000$ \\
$-7$ & $0.9118$ & $0.5000$ & $1.0000$ & $0.5000$ & $1.0000$ & $0.0000$ & $1.0000$ \\
$-6$ & $0.8921$ & $0.5000$ & $1.0000$ & $0.5000$ & $1.0000$ & $0.0001$ & $0.9996$ \\
$-5$ & $0.8689$ & $0.4999$ & $1.0000$ & $0.4999$ & $1.0000$ & $0.0003$ & $0.9988$ \\
$-4$ & $0.8418$ & $0.4997$ & $1.0000$ & $0.4997$ & $1.0000$ & $0.0006$ & $0.9976$ \\
$-3$ & $0.8106$ & $0.4993$ & $1.0000$ & $0.4993$ & $1.0000$ & $0.0013$ & $0.9948$ \\
$-2$ & $0.7753$ & $0.4981$ & $1.0000$ & $0.4981$ & $1.0000$ & $0.0029$ & $0.9884$ \\
$-1$ & $0.7360$ & $0.4953$ & $0.9999$ & $0.4953$ & $0.9999$ & $0.0060$ & $0.9759$ \\
$0$ & $0.6931$ & $0.4890$ & $0.9995$ & $0.4890$ & $0.9995$ & $0.0117$ & $0.9527$ \\
$1$ & $0.6473$ & $0.4760$ & $0.9977$ & $0.4760$ & $0.9977$ & $0.0215$ & $0.9118$ \\
$2$ & $0.5992$ & $0.4515$ & $0.9906$ & $0.4515$ & $0.9906$ & $0.0363$ & $0.8452$ \\
$3$ & $0.5498$ & $0.4100$ & $0.9676$ & $0.4100$ & $0.9676$ & $0.0553$ & $0.7464$ \\
$4$ & $0.5001$ & $0.3493$ & $0.9091$ & $0.3493$ & $0.9091$ & $0.0731$ & $0.6168$ \\
$5$ & $0.4510$ & $0.2717$ & $0.7915$ & $0.2717$ & $0.7915$ & $0.0814$ & $0.4659$ \\
$6$ & $0.4033$ & $0.1878$ & $0.6101$ & $0.1878$ & $0.6101$ & $0.0740$ & $0.3139$ \\
$7$ & $0.3579$ & $0.1126$ & $0.3998$ & $0.1126$ & $0.3998$ & $0.0538$ & $0.1847$ \\
$8$ & $0.3153$ & $0.0569$ & $0.2145$ & $0.0569$ & $0.2145$ & $0.0306$ & $0.0920$ \\
$9$ & $0.2758$ & $0.0237$ & $0.0925$ & $0.0237$ & $0.0925$ & $0.0136$ & $0.0381$ \\
$10$ & $0.2398$ & $0.0078$ & $0.0308$ & $0.0078$ & $0.0308$ & $0.0046$ & $0.0125$
\end{tabular}
\end{center}
\end{table}
\begin{figure}[tb]
\begin{center}
\includegraphics[width=10.0cm,clip]{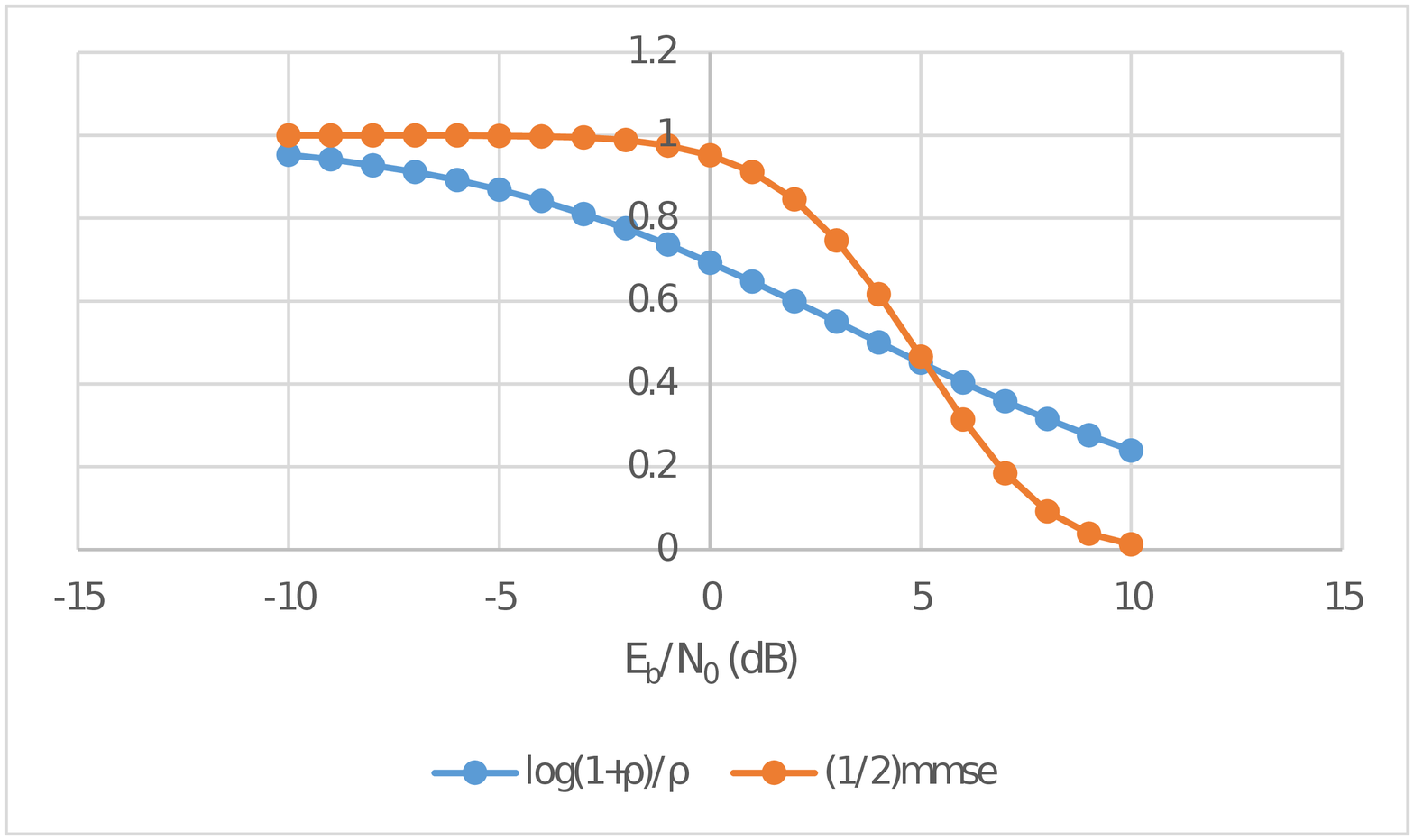}
\end{center}
\caption{$\frac{\log(1+\rho)}{\rho}$ and $\frac{1}{2}\mbox{mmse}=\frac{1}{2}\xi(\alpha_1, \alpha_2, \delta)~(C_2)$.}
\label{Fig.2}
\end{figure}
\par
With respect to the behaviors of $\frac{\log(1+\rho)}{\rho}$ and $\frac{1}{2}\xi(\alpha_1, \alpha_2, \delta)$, we have the following:
\par
(i) $\frac{\log(1+\rho)}{\rho}$:
\begin{itemize}
\item [1)] $\epsilon \rightarrow 1/2$: Since $\rho \rightarrow 0$, $\frac{\log(1+\rho)}{\rho} \rightarrow 1$ .
\item [2)] $\epsilon \rightarrow 0$: Since $\rho \rightarrow \infty$, $\frac{\log(1+\rho)}{\rho} \rightarrow 0$ .
\end{itemize}
Note that $\frac{\log(1+\rho)}{\rho}$ approaches $0$ slowly as the SNR increases.
\par
(ii) $\frac{1}{2}\xi(\alpha_1, \alpha_2, \delta)$:
\begin{itemize}
\item [1)] $\epsilon \rightarrow 1/2$: Since $\alpha_i \rightarrow 1/2~(i=1, 2)$ and since $\delta \rightarrow 0$,
\begin{equation}
\frac{1}{2}\xi(\alpha_1, \alpha_2, \delta)=\frac{1}{2}(4\alpha_1(1-\alpha_1)+4\alpha_2(1-\alpha_2)-8\delta) \rightarrow 1 .
\end{equation}
\item [2)] $\epsilon \rightarrow 0$: Since $\alpha_i \rightarrow 0~(i=1, 2)$ and since $\delta \rightarrow 0$,
\begin{equation}
\frac{1}{2}\xi(\alpha_1, \alpha_2, \delta)=\frac{1}{2}(4\alpha_1(1-\alpha_1)+4\alpha_2(1-\alpha_2)-8\delta) \rightarrow 0 .
\end{equation}
\end{itemize}
\par
From Figs. 1 and 2, we observe that the behaviors of $\frac{1}{2}\xi(\alpha_1, \alpha_2, \delta)$ for $C_1$ and $C_2$ are considerably different. Let us examine the cause of the difference.
\par
Let $\mbox{\boldmath $v$}_k=(v_k^{(1)}, v_k^{(2)})$ be the encoded block for the main decoder. We have already shown that the degree of correlation between $v_k^{(1)}$ and $v_k^{(2)}$ is expressed in terms of $\delta$ (see Lemma 3). We also see that the degree of correlation between $v_k^{(1)}$ and $v_k^{(2)}$ has a close connection with the number of error terms (denoted $m_v$) by which $v_k^{(1)}$ and $v_k^{(2)}$ differ. In fact, when $m_v$ is small, $v_k^{(1)}$ and $v_k^{(2)}$ have a strong correlation, whereas when $m_v$ is large, $v_k^{(1)}$ and $v_k^{(2)}$ have a weak correlation. These observations show that $\delta$ is closely related to $m_v$. Then it is expected that the behavior of $\frac{1}{2}\xi(\alpha_1, \alpha_2, \delta)$ depends heavily on the value of $m_v$ (i.e., on a code). In order to confirm this fact, let us evaluate the values of $m_v$ for $C_1$ and $C_2$. Note that since the encoded block is expressed as $\mbox{\boldmath $v$}_k=(\mbox{\boldmath $e$}_kG^{-1})G$, $m_v$ is obtained from $G^{-1}G$.
\par
$C_1$: The inverse encoder is given by
\begin{equation}
G_1^{-1}=\left(
\begin{array}{c}
D  \\
1+D 
\end{array}
\right) .
\end{equation}
Hence from
\begin{equation}
G_1^{-1}G_1=\left(
\begin{array}{cc}
D+\mbox{\boldmath $D$}^2+D^3 & D+D^3 \\
1+D^3 & 1+\mbox{\boldmath $D$}+\mbox{\boldmath $D$}^2+D^3
\end{array}
\right) ,
\end{equation}
it follows that $m_v=3$ (i.e., bold-faced terms).
\par
$C_2$: The inverse encoder is given by
\begin{equation}
G_2^{-1}=\left(
\begin{array}{c}
D^3+D^4+D^5  \\
1+D+D^3+D^4+D^5 
\end{array}
\right) .
\end{equation}
Then it is shown that
\begin{equation}
G_2^{-1}G_2=\left(
\begin{array}{cc}
D^3+D^{10}+D^{11} & D^3+\mbox{\boldmath $D$}^5+\mbox{\boldmath $D$}^6+\mbox{\boldmath $D$}^7+D^{10}+D^{11} \\
1+\mbox{\boldmath $D$}^2+\mbox{\boldmath $D$}^5+\mbox{\boldmath $D$}^6+\mbox{\boldmath $D$}^7+D^{10}+D^{11} & 1+\mbox{\boldmath $D$}^3+D^{10}+D^{11}
\end{array}
\right) .
\end{equation}
We see that $m_v=8$.
\par
Now compare the values of $\delta$ in Tables I and II. We observe that they are considerably different. For example, at the SNR of $E_b/N_0=0\,\mbox{dB}$, we have
\begin{equation}
\delta~(C_1)=0.0758>\delta~(C_2)=0.0117.
\end{equation}
We think the difference is due to the fact
\begin{equation}
m_v (C_1)=3<m_v (C_2)=8 .
\end{equation}
$m_v(C_1)=3$ means that there is a certain correlation between $v_k^{(1)}$ and $v_k^{(2)}$ and then we have $\delta=0.0758$ (i.e., $8\delta=0.6064$). This value is fairly large. On the other hand, $m_v(C_2)=8$ implies that $v_k^{(1)}$ and $v_k^{(2)}$ have a weak correlation, which results in a small value of $\delta=0.0117$.
\par
We have already seen that the behavior of $\frac{1}{2}\xi(\alpha_1, \alpha_2, \delta)$ is dependent on $m_v$. Also, it has been shown that the values of $m_v$ for $C_1$ and $C_2$ are considerably different. We think these explain the behaviors of curves in Figs. 1 and 2.
\par
As observed above, since the behavior of $\frac{1}{2}\xi(\alpha_1, \alpha_2, \delta)$ varies with the codes, it is appropriate to take the average with respect to a code. Then including $C_1$ and $C_2$, we have taken additionally two QLI codes $C_3$ and $C_4$ whose generator matrices are defined by
\begin{equation}
G_3(D)=(1+D+D^2+D^3, 1+D+D^3)
\end{equation}
and
\begin{equation}
G_4(D)=(1+D^3+D^4, 1+D+D^3+D^4) ,
\end{equation}
respectively. It is shown that $C_3$ and $C_4$ have $m_v=4$ and $m_v=5$, respectively. Then we have averaged the corresponding $\frac{1}{2}\xi(\alpha_1, \alpha_2, \delta)$'s. The result is shown in Fig.3, where $\frac{1}{2}\mbox{mmse(av)}$ denotes the average value of $\frac{1}{2}\xi(\alpha_1, \alpha_2, \delta)$ over four codes. We think the relation between $\frac{\log(1+\rho)}{\rho}$ and $\frac{1}{2}\xi(\alpha_1, \alpha_2, \delta)$ is shown more properly in Fig.3.
\begin{figure}[tb]
\begin{center}
\includegraphics[width=10.0cm,clip]{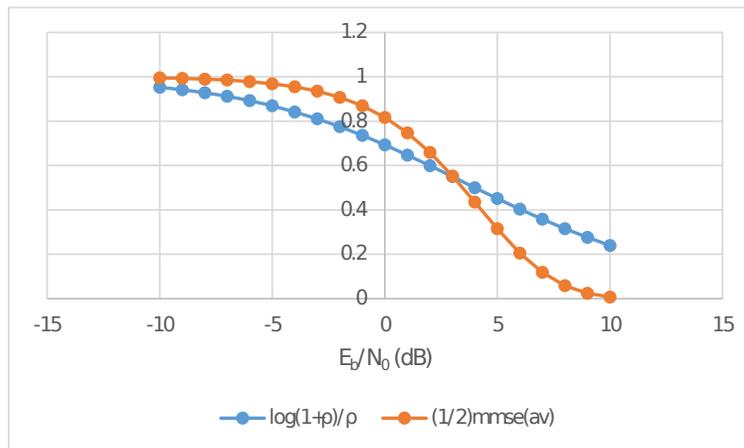}
\end{center}
\caption{$\frac{\log(1+\rho)}{\rho}$ and the average of $\frac{1}{2}\mbox{mmse}$'s.}
\label{Fig.3}
\end{figure}
\par
In Figs. 1, 2, and 3, we observe that $\frac{1}{2}\xi(\alpha_1, \alpha_2, \delta) > \frac{\log(1+\rho)}{\rho}$ holds at low-to-medium SNRs. However, the sign of inequality is reversed at high SNRs. We think this comes from the fact that $\frac{1}{2}\xi(\alpha_1, \alpha_2, \delta)$ approaches $0$ rapidly as the SNR increases, whereas $\frac{\log(1+\rho)}{\rho}$ approaches $0$ slowly as the SNR increases.

% end of file

\subsection{QLI Codes}
In~\cite{taji 19}, we have shown that the innovations approach to Viterbi decoding of QLI codes is related to ``smoothing'' in the linear estimation theory. As a result, the relationship between the mutual information and the MMSE can be discussed using the result in~\cite[Corollary 3]{guo 05}. For the purpose, we provide a proposition.
\begin{pro} Let $\mbox{\boldmath $x$}=(x_1, \cdots, x_n)$, where $x_i$ is the equiprobable binary input with unit variance. Each $x_i$ is assumed to be independent of all others. Also, let $\mbox{\boldmath $\tilde x$}=(\tilde x_1, \cdots, \tilde x_n)$ be a standard Gaussian vector. Moreover, let {\boldmath $w$} be a standard Gaussian vector which is independent of {\boldmath $x$} and $\mbox{\boldmath $\tilde x$}$. Then we have
\begin{equation}
\frac{d}{d \rho}I[\mbox{\boldmath $x$} ; \sqrt{\rho}\,\mbox{\boldmath $x$}+\mbox{\boldmath $w$}] \leq \frac{d}{d \rho}I[\mbox{\boldmath $\tilde x$} ; \sqrt{\rho}\,\mbox{\boldmath $\tilde x$}+\mbox{\boldmath $w$}] .
\end{equation}
\end{pro}
\begin{IEEEproof}
We follow Guo et al.~\cite{guo 05}. Let $\mbox{\boldmath $z$}=\sqrt{\rho}\,\mbox{\boldmath $x$}+\mbox{\boldmath $w$}$, where $\mbox{\boldmath $z$}=(z_1, \cdots, z_n)$. We have
\begin{eqnarray}
\mbox{mmse}(\rho) &=& E[(\mbox{\boldmath $x$}-E[\mbox{\boldmath $x$} \vert \mbox{\boldmath $z$}])(\mbox{\boldmath $x$}-E[\mbox{\boldmath $x$} \vert \mbox{\boldmath $z$}])^T] \nonumber \\
&=& E[\vert x_1-E[x_1 \vert \mbox{\boldmath $z$}]\vert ^2]+\cdots +E[\vert x_n-E[x_n \vert \mbox{\boldmath $z$}]  \vert ^2] \nonumber \\
&\leq& E[\vert x_1-E[x_1 \vert z_1]\vert ^2]+\cdots +E[\vert x_n-E[x_n \vert z_n]\vert ^2] \nonumber \\
&=& \sum_{i=1}^n\mbox{mmse}_i(\rho) ,
\end{eqnarray}
where $\mbox{mmse}_i(\rho)\stackrel{\triangle}{=}E[\vert x_i-E[x_i \vert z_i]\vert ^2]$. Then it follows from~\cite[Theorems 1 and 2]{guo 05} that
\begin{eqnarray}
\frac{d}{d \rho}I[\mbox{\boldmath $x$} ; \sqrt{\rho}\,\mbox{\boldmath $x$}+\mbox{\boldmath $w$}] &=& \frac{1}{2}\mbox{mmse}(\rho) \nonumber \\
&\leq& \sum_{i=1}^n\frac{1}{2}\mbox{mmse}_i(\rho) \nonumber \\
&=& \sum_{i=1}^n\frac{d}{d \rho}I[x_i; \sqrt{\rho}\,x_i+w_i] .
\end{eqnarray}
\par
({\it Remark 1:} This relation holds for any random vector {\boldmath $x$} satisfying $E[\mbox{\boldmath $x$}\mbox{\boldmath $x$}^T ]<\infty$ (cf.~\cite{guo 05}).)
\par
On the other hand, since $\tilde x_i$ is Gaussian (see~\cite[Section 9.4]{cov 06}), we have
\begin{equation}
\frac{d}{d \rho}I[\mbox{\boldmath $\tilde x$} ; \sqrt{\rho}\,\mbox{\boldmath $\tilde x$}+\mbox{\boldmath $w$}]=\sum_{i=1}^n\frac{d}{d \rho}I[\tilde x_i; \sqrt{\rho}\,\tilde x_i+w_i] .
\end{equation}
Note that in the scalar case,
\begin{equation}
\frac{d}{d \rho}I[x_i; \sqrt{\rho}\,x_i+w_i] \leq \frac{d}{d \rho}I[\tilde x_i; \sqrt{\rho}\,\tilde x_i+w_i]
\end{equation}
holds (see Appendix D and see~\cite[Fig.1]{guo 05}). Hence we have
\begin{eqnarray}
\frac{d}{d \rho}I[\mbox{\boldmath $x$} ; \sqrt{\rho}\,\mbox{\boldmath $x$}+\mbox{\boldmath $w$}] &\leq& \sum_{i=1}^n\frac{d}{d \rho}I[x_i; \sqrt{\rho}\,x_i+w_i] \nonumber \\
&\leq& \sum_{i=1}^n\frac{d}{d \rho}I[\tilde x_i; \sqrt{\rho}\,\tilde x_i+w_i] \nonumber \\
&=& \frac{d}{d \rho}I[\mbox{\boldmath $\tilde x$} ; \sqrt{\rho}\,\mbox{\boldmath $\tilde x$}+\mbox{\boldmath $w$}] . \nonumber
\end{eqnarray}
\end{IEEEproof}
\par
Now the following~\cite[Corollary 3]{guo 05} has been shown:
\begin{equation}
\frac{d}{d\rho}I[\mbox{\boldmath $x$}^k; \mbox{\boldmath $z$}^k]=\frac{1}{2}\sum_{i=1}^k \mbox{mmse}(i, \rho) ,
\end{equation}
where
\begin{equation}
\mbox{mmse}(i, \rho)\stackrel{\triangle}{=}E[(\mbox{\boldmath $x$}_i-E[\mbox{\boldmath $x$}_i \vert \mbox{\boldmath $z$}^k])(\mbox{\boldmath $x$}_i-E[\mbox{\boldmath $x$}_i \vert \mbox{\boldmath $z$}^k])^T]~(1 \leq i \leq k) .
\end{equation}
\par
{\it Remark 2:} The definition of $\mbox{mmse}(i, \rho)$ is slightly different from that in~\cite{guo 05}. However, since the whole observations $\mbox{\boldmath $z$}^k$ are used in each conditional expectation, the above equality holds also under our definition.
\par
First note the left-hand side, i.e., $\frac{d}{d\rho}I[\mbox{\boldmath $x$}^k; \mbox{\boldmath $z$}^k]$. Let $\mbox{\boldmath $\tilde x$}_k$ be Gaussian. It follows from Proposition 10 that
\begin{equation}
\frac{d}{d\rho}I[\mbox{\boldmath $\tilde x$}^k; \mbox{\boldmath $z$}^k] \geq \frac{d}{d\rho}I[\mbox{\boldmath $x$}^k; \mbox{\boldmath $z$}^k] .
\end{equation}
Since $\mbox{\boldmath $\tilde x$}_k$ is Gaussian, we have
\begin{equation}
I[\mbox{\boldmath $\tilde x$}^k; \mbox{\boldmath $z$}^k]=\frac{1}{2}\log(1+\rho)^{2k}=k\log(1+\rho)
\end{equation}
and hence
\begin{equation}
\frac{d}{d\rho}I[\mbox{\boldmath $\tilde x$}^k; \mbox{\boldmath $z$}^k]=\frac{k}{1+\rho}
\end{equation}
holds. Therefore,
\begin{equation}
\frac{1}{1+\rho} \geq \frac{d}{d\rho}\left(\frac{I[\mbox{\boldmath $x$}^k; \mbox{\boldmath $z$}^k]}{k}\right)=\frac{1}{2}\left(\frac{1}{k}\sum_{i=1}^k \mbox{mmse}(i, \rho)\right)
\end{equation}
is obtained.
\par
Next, note the right-hand side, i.e., $\frac{1}{2}\sum_{i=1}^k \mbox{mmse}(i, \rho)$. For $1+L \leq k$, we have
\begin{eqnarray}
\lefteqn{\sum_{i=1}^k \mbox{mmse}(i, \rho)} \nonumber \\
&& =E[(\mbox{\boldmath $x$}_1-E[\mbox{\boldmath $x$}_1 \vert \mbox{\boldmath $z$}^k])(\mbox{\boldmath $x$}_1-E[\mbox{\boldmath $x$}_1 \vert \mbox{\boldmath $z$}^k])^T] \nonumber \\
&& \cdots \nonumber \\
&& +E[(\mbox{\boldmath $x$}_{k-L}-E[\mbox{\boldmath $x$}_{k-L} \vert \mbox{\boldmath $z$}^k])(\mbox{\boldmath $x$}_{k-L}-E[\mbox{\boldmath $x$}_{k-L} \vert \mbox{\boldmath $z$}^k])^T] \nonumber \\
&& \cdots \nonumber \\
&& +E[(\mbox{\boldmath $x$}_k-E[\mbox{\boldmath $x$}_k \vert \mbox{\boldmath $z$}^k])(\mbox{\boldmath $x$}_k-E[\mbox{\boldmath $x$}_k \vert \mbox{\boldmath $z$}^k])^T] .
\end{eqnarray}
Our concern is the evaluation of
\begin{eqnarray}
\lefteqn{\mbox{mmse}(k-L, \rho)} \nonumber \\
&& =E[(\mbox{\boldmath $x$}_{k-L}-E[\mbox{\boldmath $x$}_{k-L} \vert \mbox{\boldmath $z$}^k])(\mbox{\boldmath $x$}_{k-L}-E[\mbox{\boldmath $x$}_{k-L} \vert \mbox{\boldmath $z$}^k])^T] .
\end{eqnarray}
For the purpose, we assume an approximation:
\begin{equation}
\mbox{mmse}(k-L, \rho) \approx \frac{1}{k}\sum_{i=1}^k \mbox{mmse}(i, \rho) .
\end{equation}
Hence in the relation
\begin{eqnarray}
\frac{1}{1+\rho} &\geq& \frac{d}{d\rho}\left(\frac{I[\mbox{\boldmath $x$}^k; \mbox{\boldmath $z$}^k]}{k}\right) \nonumber \\
&=& \frac{1}{2}\left(\frac{1}{k}\sum_{i=1}^k \mbox{mmse}(i, \rho)\right) \nonumber \\
&\approx& \frac{1}{2}\mbox{mmse}(k-L, \rho) ,
\end{eqnarray}
we replace $\mbox{mmse}(k-L, \rho)$ by $\xi(\beta_1, \beta_2, \delta')$. Then we have
\begin{equation}
\frac{1}{1+\rho} \geq \frac{d}{d\rho}\left(\frac{I[\mbox{\boldmath $x$}^k; \mbox{\boldmath $z$}^k]}{k}\right)\approx \frac{1}{2}\xi(\beta_1, \beta_2, \delta') .
\end{equation}
Note that the above approximation is equivalent to set
\begin{equation}
\frac{1}{k}\sum_{i=1}^k \mbox{mmse}(i, \rho) \approx \xi(\beta_1, \beta_2, \delta') ,
\end{equation}
where the left-hand side represents the average noncausal MMSE (see~\cite[Section IV-A]{guo 05}).

\subsection{Numerical Results as QLI Codes}
In order to confirm the validity of the derived equation, numerical calculations have been carried out. We have used the same QLI codes $C_1$ and $C_2$ as in the previous section. In this case, we regard these codes as inherent QLI codes and compare the values of $\frac{1}{1+\rho}$ and $\frac{1}{2}\xi(\beta_1, \beta_2, \delta')$. The results are shown in Tables III and IV, where $\frac{1}{2}\xi(\beta_1, \beta_2, \delta')$ is simply denoted $\frac{1}{2}\mbox{mmse}$. The results for $C_1$ and $C_2$ are shown also in Fig.4 and Fig.5, respectively.
\begin{table}[tb]
\caption{$\frac{1}{1+\rho}$ and minimum mean-square error ($C_1$ as a QLI code)}
\label{Table 3}
\begin{center}
\begin{tabular}{c*{7}{|c}}
$E_b/N_0~(\mbox{dB})$ & $\frac{1}{1+\rho}$ & $\beta_1$ & $4\beta_1(1-\beta_1)$ & $\beta_2$ & $4\beta_2(1-\beta_2)$ & $\delta'$ & $\frac{1}{2}\mbox{mmse}$ \\
\hline
$-10$ & $0.9091$ & $0.4999$ & $1.0000$ & $0.4981$ & $1.0000$ & $0.0154$ & $0.9384$ \\
$-9$ & $0.8882$ & $0.4998$ & $1.0000$ & $0.4970$ & $1.0000$ & $0.0192$ & $0.9232$ \\
$-8$ & $0.8632$ & $0.4996$ & $1.0000$ & $0.4954$ & $0.9999$ & $0.0239$ & $0.9044$ \\
$-7$ & $0.8337$ & $0.4992$ & $1.0000$ & $0.4929$ & $0.9998$ & $0.0297$ & $0.8811$ \\
$-6$ & $0.7992$ & $0.4984$ & $1.0000$ & $0.4892$ & $0.9995$ & $0.0368$ & $0.8526$ \\
$-5$ & $0.7597$ & $0.4970$ & $1.0000$ & $0.4835$ & $0.9989$ & $0.0453$ & $0.8183$ \\
$-4$ & $0.7153$ & $0.4945$ & $0.9999$ & $0.4752$ & $0.9975$ & $0.0555$ & $0.7767$ \\
$-3$ & $0.6661$ & $0.4900$ & $0.9996$ & $0.4632$ & $0.9946$ & $0.0674$ & $0.7275$ \\
$-2$ & $0.6131$ & $0.4823$ & $0.9987$ & $0.4461$ & $0.9884$ & $0.0811$ & $0.6692$ \\
$-1$ & $0.5573$ & $0.4696$ & $0.9963$ & $0.4226$ & $0.9760$ & $0.0959$ & $0.6026$ \\
$0$ & $0.5000$ & $0.4494$ & $0.9898$ & $0.3914$ & $0.9528$ & $0.1110$ & $0.5273$ \\
$1$ & $0.4427$ & $0.4191$ & $0.9738$ & $0.3515$ & $0.9118$ & $0.1242$ & $0.4460$ \\
$2$ & $0.3869$ & $0.3766$ & $0.9391$ & $0.3033$ & $0.8452$ & $0.1326$ & $0.3618$ \\
$3$ & $0.3339$ & $0.3213$ & $0.8723$ & $0.2482$ & $0.7464$ & $0.1325$ & $0.2794$ \\
$4$ & $0.2847$ & $0.2565$ & $0.7628$ & $0.1905$ & $0.6168$ & $0.1213$ & $0.2046$ \\
$5$ & $0.2403$ & $0.1876$ & $0.6096$ & $0.1346$ & $0.4659$ & $0.0995$ & $0.1398$ \\
$6$ & $0.2008$ & $0.1231$ & $0.4318$ & $0.0858$ & $0.3138$ & $0.0714$ & $0.0872$ \\
$7$ & $0.1663$ & $0.0710$ & $0.2638$ & $0.0485$ & $0.1846$ & $0.0439$ & $0.0486$ \\
$8$ & $0.1368$ & $0.0349$ & $0.1347$ & $0.0236$ & $0.0922$ & $0.0225$ & $0.0235$ \\
$9$ & $0.1118$ & $0.0143$ & $0.0564$ & $0.0096$ & $0.0380$ & $0.0095$ & $0.0092$ \\
$10$ & $0.0909$ & $0.0047$ & $0.0187$ & $0.0031$ & $0.0124$ & $0.0031$ & $0.0032$
\end{tabular}
\end{center}
\end{table}
\begin{figure}[tb]
\begin{center}
\includegraphics[width=10.0cm,clip]{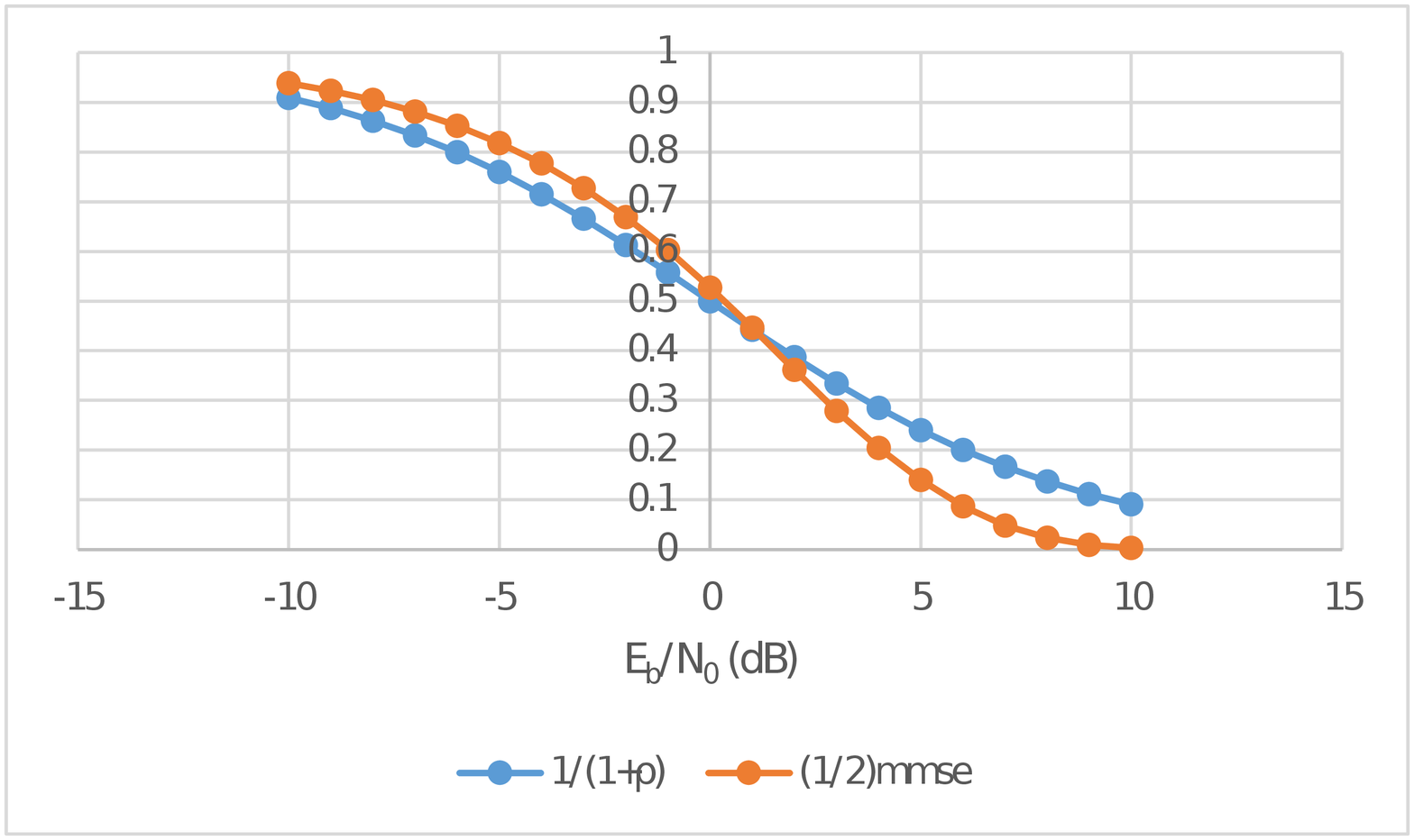}
\end{center}
\caption{$\frac{1}{1+\rho}$ and $\frac{1}{2}\mbox{mmse}=\frac{1}{2}\xi(\beta_1, \beta_2, \delta')~(C_1)$.}
\label{Fig.4}
\end{figure}
\begin{table}[tb]
\caption{$\frac{1}{1+\rho}$ and minimum mean-square error ($C_2$ as a QLI code)}
\label{Table 4}
\begin{center}
\begin{tabular}{c*{7}{|c}}
$E_b/N_0~(\mbox{dB})$ & $\frac{1}{1+\rho}$ & $\beta_1$ & $4\beta_1(1-\beta_1)$ & $\beta_2$ & $4\beta_2(1-\beta_2)$ & $\delta'$ & $\frac{1}{2}\mbox{mmse}$ \\
\hline
$-10$ & $0.9091$ & $0.5000$ & $1.0000$ & $0.5000$ & $1.0000$ & $0.0154$ & $0.9384$ \\
$-9$ & $0.8882$ & $0.5000$ & $1.0000$ & $0.5000$ & $1.0000$ & $0.0192$ & $0.9232$ \\
$-8$ & $0.8632$ & $0.5000$ & $1.0000$ & $0.5000$ & $1.0000$ & $0.0239$ & $0.9044$ \\
$-7$ & $0.8337$ & $0.5000$ & $1.0000$ & $0.5000$ & $1.0000$ & $0.0297$ & $0.8812$ \\
$-6$ & $0.7992$ & $0.5000$ & $1.0000$ & $0.5000$ & $1.0000$ & $0.0368$ & $0.8528$ \\
$-5$ & $0.7597$ & $0.4999$ & $1.0000$ & $0.5000$ & $1.0000$ & $0.0454$ & $0.8184$ \\
$-4$ & $0.7153$ & $0.4997$ & $1.0000$ & $0.4999$ & $1.0000$ & $0.0557$ & $0.7772$ \\
$-3$ & $0.6661$ & $0.4993$ & $1.0000$ & $0.4998$ & $1.0000$ & $0.0678$ & $0.7288$ \\
$-2$ & $0.6131$ & $0.4981$ & $1.0000$ & $0.4994$ & $1.0000$ & $0.0820$ & $0.6720$ \\
$-1$ & $0.5573$ & $0.4953$ & $0.9999$ & $0.4981$ & $1.0000$ & $0.0984$ & $0.6064$ \\
$0$ & $0.5000$ & $0.4890$ & $0.9995$ & $0.4949$ & $0.9999$ & $0.1164$ & $0.5341$ \\
$1$ & $0.4427$ & $0.4760$ & $0.9977$ & $0.4869$ & $0.9993$ & $0.1359$ & $0.4548$ \\
$2$ & $0.3869$ & $0.4515$ & $0.9906$ & $0.4695$ & $0.9963$ & $0.1553$ & $0.3721$ \\
$3$ & $0.3339$ & $0.4100$ & $0.9676$ & $0.4362$ & $0.9837$ & $0.1717$ & $0.2890$ \\
$4$ & $0.2847$ & $0.3493$ & $0.9091$ & $0.3814$ & $0.9437$ & $0.1788$ & $0.2112$ \\
$5$ & $0.2403$ & $0.2717$ & $0.7915$ & $0.3048$ & $0.8476$ & $0.1692$ & $0.1429$ \\
$6$ & $0.2008$ & $0.1878$ & $0.6101$ & $0.2159$ & $0.6770$ & $0.1388$ & $0.0883$ \\
$7$ & $0.1663$ & $0.1126$ & $0.3998$ & $0.1319$ & $0.4580$ & $0.0950$ & $0.0490$ \\
$8$ & $0.1368$ & $0.0569$ & $0.2145$ & $0.0674$ & $0.2515$ & $0.0524$ & $0.0236$ \\
$9$ & $0.1118$ & $0.0237$ & $0.0925$ & $0.0283$ & $0.1099$ & $0.0229$ & $0.0096$ \\
$10$ & $0.0909$ & $0.0078$ & $0.0308$ & $0.0093$ & $0.0368$ & $0.0077$ & $0.0032$
\end{tabular}
\end{center}
\end{table}
\begin{figure}[tb]
\begin{center}
\includegraphics[width=10.0cm,clip]{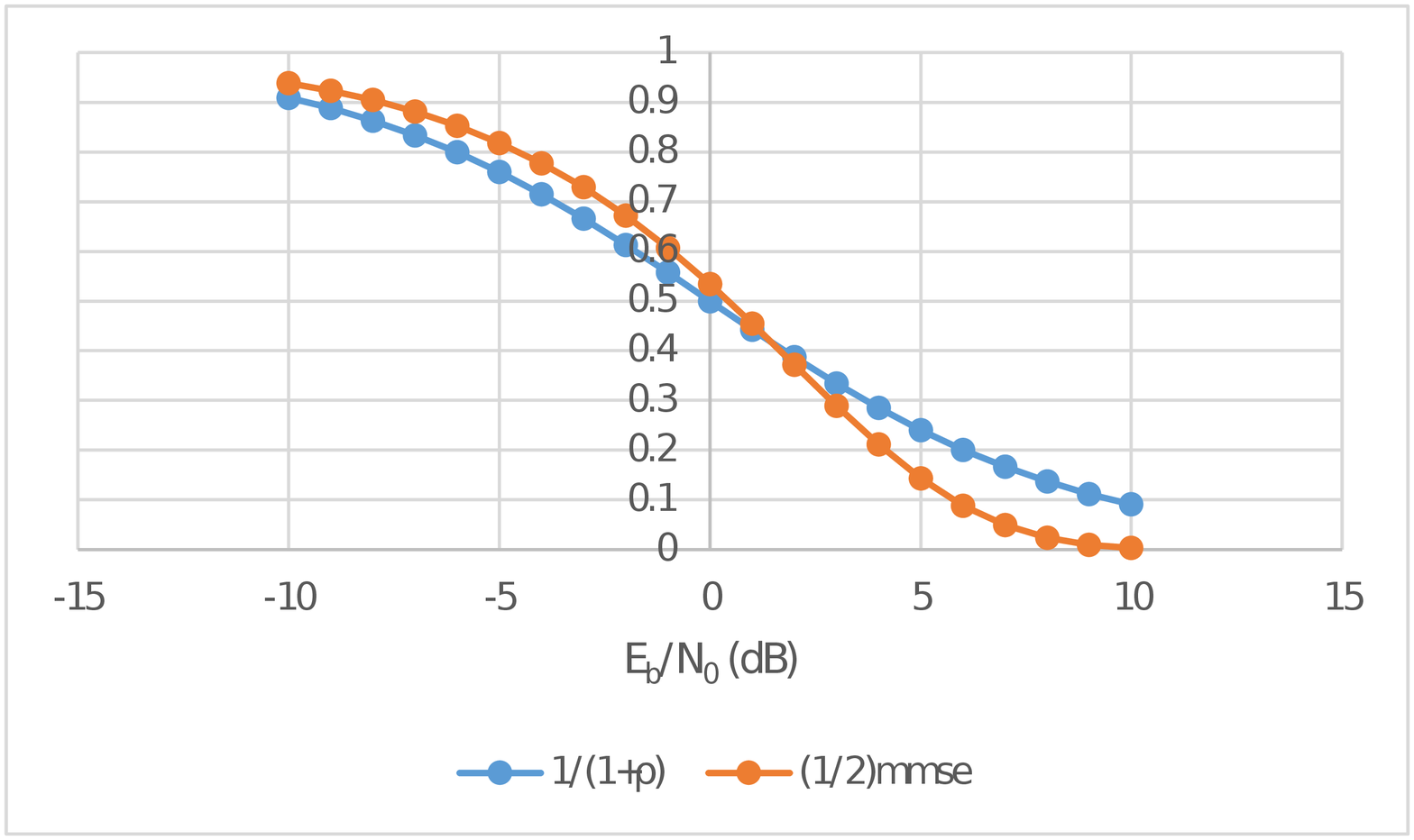}
\end{center}
\caption{$\frac{1}{1+\rho}$ and $\frac{1}{2}\mbox{mmse}=\frac{1}{2}\xi(\beta_1, \beta_2, \delta')~(C_2)$.}
\label{Fig.5}
\end{figure}
\par
First note the following.
\begin{lem}
Let $\rho>0$. Then
\begin{equation}
\frac{1}{1+\rho} \leq \frac{\log(1+\rho)}{\rho}
\end{equation}
holds.
\end{lem}
\begin{IEEEproof}
It is shown~\cite[Theorem 63]{hardy 52} that
\begin{equation}
uv \leq u\log\,u+e^{v-1}~(u>0) .
\end{equation}
Letting $v=1$, we have
\begin{displaymath}
u \leq u\log\,u+1~(u>0) .
\end{displaymath}
Furthermore, letting $u=\rho+1$, we have
\begin{displaymath}
1+\rho \leq (1+\rho)\log(1+\rho)+1~(\rho>0) .
\end{displaymath}
This is equivalent to
\begin{displaymath}
\frac{1}{1+\rho} \leq \frac{\log(1+\rho)}{\rho} .
\end{displaymath}
\end{IEEEproof}
\par
The behavior of $\frac{1}{1+\rho}$ is similar to that of $\frac{\log(1+\rho)}{\rho}$. We have the following:
\par
(i) $\frac{1}{1+\rho}$: 
\begin{itemize}
\item [1)] $\epsilon \rightarrow 1/2$: Since $\rho \rightarrow 0$, $\frac{1}{1+\rho} \rightarrow 1$.
\item [2)] $\epsilon \rightarrow 0$: Since $\rho \rightarrow \infty$, $\frac{1}{1+\rho} \rightarrow 0$.
\end{itemize}
Note that $\frac{1}{1+\rho}$ approaches $0$ more rapidly as the SNR increases compared with $\frac{\log(1+\rho)}{\rho}$.
\par
(ii) $\frac{1}{2}\xi(\beta_1, \beta_2, \delta')$:
\begin{itemize}
\item [1)] $\epsilon \rightarrow 1/2$: Since $\beta_i \rightarrow 1/2~(i=1, 2)$ and since $\delta' \rightarrow 0$,
\begin{equation}
\frac{1}{2}\xi(\beta_1, \beta_2, \delta')=\frac{1}{2}(4\beta_1(1-\beta_1)+4\beta_2(1-\beta_2)-8\delta') \rightarrow 1 .
\end{equation}
\item [2)] $\epsilon \rightarrow 0$: Since $\beta_i \rightarrow 0~(i=1, 2)$ and since $\delta' \rightarrow 0$,
\begin{equation}
\frac{1}{2}\xi(\beta_1, \beta_2, \delta')=\frac{1}{2}(4\beta_1(1-\beta_1)+4\beta_2(1-\beta_2)-8\delta') \rightarrow 0 .
\end{equation}
\end{itemize}
\par
From Tables III and IV (or Figs. 4 and 5), we observe that the behaviors of $\frac{1}{2}\xi(\beta_1, \beta_2, \delta')$ for $C_1$ and $C_2$ are almost the same. This is because both $C_1$ and $C_2$ are regarded as QLI codes and hence have the same $m_v~(=2)$, which results in almost equal values of $\delta'$. Also, observe that $\frac{1}{2}\xi(\beta_1, \beta_2, \delta')$ provides a good approximation to $\frac{1}{1+\rho}$ (Figs. 4 and 5).

% end of file

\section{Discussion}
In the previous section, we have discussed the validity of the derived MMSE using the results of Guo et al.~\cite{guo 05}. In reality, we have examined
\begin{itemize}
\item [1)] the relation between $\frac{\log(1+\rho)}{\rho}$ and $\frac{1}{2}\mbox{mmse}=\frac{1}{2}\xi(\alpha_1, \alpha_2, \delta)$,
\item [2)] the relation between $\frac{1}{1+\rho}$ and $\frac{1}{2}\mbox{mmse}=\frac{1}{2}\xi(\beta_1, \beta_2, \delta')$.
\end{itemize}
Note that the relation between the mutual information and the MMSE has been reduced to the relation between a function of $\rho$ and the MMSE. That is, their relation has been examined not directly but indirectly. Now we see that the following approximation and evaluation are used in the argument in Section IV:
\begin{itemize}
\item [i)] $\frac{1}{k}\sum_{i=1}^k\mbox{pmmse}(i, \rho) \approx \xi(\alpha_1, \alpha_2, \delta)$ ($\frac{1}{k}\sum_{i=1}^k \mbox{mmse}(i, \rho) \approx \xi(\beta_1, \beta_2, \delta')$).
\item [ii)] $I[\mbox{\boldmath $x$}^k; \mbox{\boldmath $z$}^k] \leq k\log(1+\rho)$.
\item [iii)] $\frac{d}{d\rho}I[\mbox{\boldmath $x$}^k; \mbox{\boldmath $z$}^k] \leq \frac{k}{1+\rho}$.
\end{itemize}
Since $\xi(\alpha_1, \alpha_2, \delta)$ ($\xi(\beta_1, \beta_2, \delta')$) is not dependent on $k$, averaging in i) seems to be reasonable. (But it has to be justified.) On the other hand, inequalities ii) and iii) come from the fact that the input in our signal model is not Gaussian and hence the input-output mutual information cannot be obtained in a concrete form. (In the scalar case, the mutual information of Gaussian channel with equiprobable binary input has been given (see~\cite[p.1263]{guo 05}). In our case, however, since the input is taken as a vector, we have used the conventional inequality on mutual information.) Consider the case that a QLI code is regarded as a general code. It follows from i) and ii) that
\begin{eqnarray}
\frac{1}{\rho}\left(\frac{I[\mbox{\boldmath $x$}^k; \mbox{\boldmath $z$}^k]}{k}\right) &\lesssim& \frac{1}{2}\xi(\alpha_1, \alpha_2, \delta) \nonumber \\
\frac{1}{\rho}\left(\frac{I[\mbox{\boldmath $x$}^k; \mbox{\boldmath $z$}^k]}{k}\right) &\leq& \frac{\log(1+\rho)}{\rho} . \nonumber
\end{eqnarray}
Similarly, in the case that a QLI code is regarded as the inherent QLI code, it follows from i) and iii) that
\begin{eqnarray}
\frac{d}{d\rho}\left(\frac{I[\mbox{\boldmath $x$}^k; \mbox{\boldmath $z$}^k]}{k}\right) &\approx& \frac{1}{2}\xi(\beta_1, \beta_2, \delta') \nonumber \\
\frac{d}{d\rho}\left(\frac{I[\mbox{\boldmath $x$}^k; \mbox{\boldmath $z$}^k]}{k}\right) &\leq& \frac{1}{1+\rho} . \nonumber
\end{eqnarray}
Hence evaluations of $I[\mbox{\boldmath $x$}^k; \mbox{\boldmath $z$}^k]$ and $\frac{d}{d\rho}I[\mbox{\boldmath $x$}^k; \mbox{\boldmath $z$}^k]$ are important in our discussions. It is expected that inequalities ii) and iii) are tight at low SNRs, whereas they are loose at high SNRs (cf.~\cite[p.1263]{guo 05}). As a result, if approximation i) is appropriate, then crossing of two curves in Figs. $1\sim 5$ is understandable. Hence the validity of approximation i) has to be further examined. Nevertheless, in the case that a given QLI code is regarded as the inherent QLI code, it has been shown that $\frac{1}{2}\xi(\beta_1, \beta_2, \delta')$ provides a good approximation to $\frac{1}{1+\rho}$. This is quite remarkable considering the fact that $\frac{1}{1+\rho}$ is a function of $\rho$, whereas $\frac{1}{2}\xi(\beta_1, \beta_2, \delta')$ is dependent on concrete convolutional coding. We think this fact implies the validity of our inference and the related results.
\par
Here note the following: $\frac{\log(1+\rho)}{\rho}$ ($\frac{1}{1+\rho}$) is a function only of $\rho=c^2=E_b/N_0$, whereas $\frac{1}{2}\mbox{mmse}=\frac{1}{2}\xi(\alpha_1, \alpha_2, \delta)$ ($\frac{1}{2}\xi(\beta_1, \beta_2, \delta')$) is dependent on the encoded block for the main decoder. Hence at first glance, the comparison seems to be inappropriate. Although $\frac{1}{2}\mbox{mmse}$ actually depends on coding, it is a function of the channel error probability $\epsilon$. Since $\epsilon=Q\bigl(\sqrt{2E_s/N_0}\bigr)=Q\bigl(\sqrt{E_b/N_0}\bigr)=Q\bigl(\sqrt{\rho}\bigr)$ (cf. $n_0=2$), $\frac{1}{2}\mbox{mmse}$ is also a function of $\rho$. Hence the above comparison is justified.
\par
By the way, the argument in Section IV is based on the signal model: $z_j=cx_j+w_j$ given in Section I. Note that $x_j$ is determined by the encoded symbol $y_j$ and hence this signal model is dependent clearly on convolutional coding. However, since each $y_j$ is independent of all others, $x_j$ can be seen as a random variable having values $\pm 1$ with equal probability. When $x_j$ is seen in this way, convolutional coding is not found explicitly in the expression $z_j=cx_j+w_j$. In words, we cannot see from the signal model how $\{y_j\}$ is generated. On the other hand, consider the MMSE in estimating the input based on the observations $\{z_j\}$. If the signal model is interpreted as above, then the MMSE seems to be independent of concrete convolutional coding. But this is not true. As we have already seen, the MMSE is replaced finally by $\xi(\alpha_1, \alpha_2, \delta)$ ($\xi(\beta_1, \beta_2, \delta')$), which is an essential step. By way of this replacement, the MMSE is connected with the convolutional coding. That is, convolutional coding is actually reflected in the MMSE.
\par
From the results in Section IV, it seems that our argument is more convincing for QLI codes. We know that an SST Viterbi decoder functions well for QLI codes compared with general codes. In fact, QLI codes are preferable from a likelihood concentration viewpoint~\cite{pin 91,taji 19}. Then a question arises: How is a likelihood concentration in the main decoder related to the MMSE? Suppose that the information $u_k=\mbox{\boldmath $e$}_kG^{-1}(D)$ ($u_k=\mbox{\boldmath $e$}_kF$) for the main decoder consists of $m_u$ error terms. We know that a likelihood concentration in the main decoder depends on $m_u$~\cite{taji 19}, whereas $\delta~(\delta')$ is affected by $m_v$ and hence the MMSE is dependent on $m_v$.
\par
In the case of QLI codes, we have the following.
\begin{pro}
Consider a QLI code whose generator matrix is given by
\begin{displaymath}
G(D)=(g_1(D), g_1(D)+D^L)~(1 \leq L \leq \nu-1) .
\end{displaymath}
We have
\begin{equation}
m_u=m_v .
\end{equation}
\end{pro}
\begin{IEEEproof}
Let $u_k=e_1+ \cdots +e_m$, where errors $e_j~(1 \leq j \leq m)$ are mutually independent. We have
\begin{eqnarray}
\mbox{\boldmath $v$}_k &=& u_kG(D) \nonumber \\
&=& (u_kg_1(D), u_kg_1(D)+u_kD^L) \nonumber \\
&=& (v_k^{(1)}, v_k^{(2)}) .
\end{eqnarray}
Then it follows that $v_k^{(1)}$ and $v_k^{(2)}$ differ by $u_kD^L$. Since $u_k=e_1+ \cdots +e_m$, $v_k^{(1)}$ and $v_k^{(2)}$ differ by $m$ error terms.
\end{IEEEproof}
\par
We observe that the relation $m_u=m_v$ actually holds for $C_1 \sim C_4$ (see Section IV-B). The above result shows that in the case of QLI codes, a likelihood concentration in the main decoder and the degree of correlation between $v_k^{(1)}$ and $v_k^{(2)}$ (i.e., the value of $\delta~(\delta')$) have a close connection. We remark that the former is independent of the latter in principle. The above result, however, shows that the two notions are closely related to each other.
\par
Note that the equality $m_u=m_v$ does not hold for general codes in general. Then in order to examine the relation between a likelihood concentration in the main decoder and the MMSE, we can consider a code with small $m_u$ and large $m_v$. As an example, take the code $C_5$~\cite{pin 91} whose generator matrix is defined by
\begin{equation}
G_5(D)=(1+D+D^4+D^5+D^6, 1+D^2+D^3+D^4+D^6) .
\end{equation}
The inverse encoder is given by
\begin{equation}
G_5^{-1}=\left(
\begin{array}{c}
D  \\
1+D 
\end{array}
\right)
\end{equation}
and we have $m_u=3$. On the other hand, it is shown that
\begin{equation}
G_5^{-1}G_5=\left(
\begin{array}{cc}
D+\mbox{\boldmath $D$}^2+D^5+\mbox{\boldmath $D$}^6+D^7 & D+\mbox{\boldmath $D$}^3+\mbox{\boldmath $D$}^4+D^5+D^7 \\
1+D^2+\mbox{\boldmath $D$}^4+D^7 & 1+\mbox{\boldmath $D$}+D^2+\mbox{\boldmath $D$}^5+\mbox{\boldmath $D$}^6+D^7
\end{array}
\right) .
\end{equation}
Then we have $m_v=8$.
\par
Since $m_u$ is small, a likelihood concentration occurs in the main decoder~\cite{pin 91,taji 19}. On the other hand, $m_v=8$ is considerably large. Here recall that when $C_2$ is regarded as a general code, it has $m_v=8$. This value is the same as that for $C_5$. However, since $C_2$ has $m_u=m_v=8$ as a general code, a likelihood concentration in the main decoder is not expected at low-to-medium SNRs. Then in order to see the effect of $m_u$ on the values of $\frac{1}{2}\mbox{mmse}$, let us compare the values of $\frac{1}{2}\mbox{mmse}=\frac{1}{2}\xi(\alpha_1, \alpha_2, \delta)$ for $C_2$ and $C_5$. For the purpose, we have evaluated $\frac{1}{2}\mbox{mmse}$ for $C_5$. The result is shown in Table V.
\begin{table}[tb]
\caption{$\frac{\log(1+\rho)}{\rho}$ and minimum mean-square error ($C_5$)}
\label{Table 5}
\begin{center}
\begin{tabular}{c*{7}{|c}}
$E_b/N_0~(\mbox{dB})$ & $\frac{\log(1+\rho)}{\rho}$ & $\alpha_1$ & $4\alpha_1(1-\alpha_1)$ & $\alpha_2$ & $4\alpha_2(1-\alpha_2)$ & $\delta$ & $\frac{1}{2}\mbox{mmse}$ \\
\hline
$-10$ & $0.9531$ & $0.5000$ & $1.0000$ & $0.5000$ & $1.0000$ & $0.0000$ & $1.0000$ \\
$-9$ & $0.9419$ & $0.5000$ & $1.0000$ & $0.5000$ & $1.0000$ & $0.0000$ & $1.0000$ \\
$-8$ & $0.9282$ & $0.5000$ & $1.0000$ & $0.5000$ & $1.0000$ & $0.0000$ & $1.0000$ \\
$-7$ & $0.9118$ & $0.5000$ & $1.0000$ & $0.5000$ & $1.0000$ & $0.0000$ & $1.0000$ \\
$-6$ & $0.8921$ & $0.4999$ & $1.0000$ & $0.5000$ & $1.0000$ & $0.0002$ & $0.9992$ \\
$-5$ & $0.8689$ & $0.4998$ & $1.0000$ & $0.5000$ & $1.0000$ & $0.0002$ & $0.9992$ \\
$-4$ & $0.8418$ & $0.4994$ & $1.0000$ & $0.4999$ & $1.0000$ & $0.0006$ & $0.9976$ \\
$-3$ & $0.8106$ & $0.4986$ & $1.0000$ & $0.4996$ & $1.0000$ & $0.0014$ & $0.9944$ \\
$-2$ & $0.7753$ & $0.4967$ & $1.0000$ & $0.4989$ & $1.0000$ & $0.0029$ & $0.9884$ \\
$-1$ & $0.7360$ & $0.4925$ & $0.9998$ & $0.4970$ & $1.0000$ & $0.0060$ & $0.9759$ \\
$0$ & $0.6931$ & $0.4839$ & $0.9990$ & $0.4925$ & $0.9998$ & $0.0117$ & $0.9526$ \\
$1$ & $0.6473$ & $0.4675$ & $0.9958$ & $0.4823$ & $0.9987$ & $0.0214$ & $0.9117$ \\
$2$ & $0.5992$ & $0.4387$ & $0.9850$ & $0.4615$ & $0.9941$ & $0.0363$ & $0.8444$ \\
$3$ & $0.5498$ & $0.3932$ & $0.9544$ & $0.4242$ & $0.9770$ & $0.0553$ & $0.7445$ \\
$4$ & $0.5001$ & $0.3301$ & $0.8845$ & $0.3663$ & $0.9285$ & $0.0731$ & $0.6141$ \\
$5$ & $0.4510$ & $0.2531$ & $0.7562$ & $0.2889$ & $0.8217$ & $0.0814$ & $0.4634$ \\
$6$ & $0.4033$ & $0.1727$ & $0.5715$ & $0.2021$ & $0.6450$ & $0.0741$ & $0.3119$ \\
$7$ & $0.3579$ & $0.1026$ & $0.3683$ & $0.1224$ & $0.4297$ & $0.0537$ & $0.1842$ \\
$8$ & $0.3153$ & $0.0515$ & $0.1954$ & $0.0622$ & $0.2333$ & $0.0306$ & $0.0920$ \\
$9$ & $0.2758$ & $0.0214$ & $0.0838$ & $0.0260$ & $0.1013$ & $0.0136$ & $0.0382$ \\
$10$ & $0.2398$ & $0.0070$ & $0.0278$ & $0.0085$ & $0.0337$ & $0.0045$ & $0.0128$
\end{tabular}
\end{center}
\end{table}
\par
In Table V, look at the values of $\frac{1}{2}\mbox{mmse}$. We observe that they are almost the same as those for $C_2$ (see Table II). That is, it seems that $m_u$ has little effect on the values of $\frac{1}{2}\mbox{mmse}$. This observation is explained as follows. $m_u=3$ means that a likelihood concentration in the main decoder is notable. However, this does not necessarily affect the MMSE. A likelihood concentration really reduces the decoding complexity (i.e., the complexity in estimating the input). On the other hand, the MMSE is the estimation error which is finally attained after the estimation process. In other words, $m_u$ affects the complexity of estimation, whereas $m_v$ is related to the final estimation error. Moreover, we can say it as follows: the MMSE is determined by the structure of $G^{-1}G$. Hence even if $G^{-1}$'s are considerably different between two codes $C$ and $\tilde C$, when $G^{-1}G$'s have almost equal values of $m_v$, the MMSE's are close to each other. In fact, the inverse encoder for $C_2$ is given by
\begin{displaymath}
G_2^{-1}=\left(
\begin{array}{c}
D^3+D^4+D^5  \\
1+D+D^3+D^4+D^5 
\end{array}
\right) ,
\end{displaymath}
whereas the inverse encoder for $C_5$ is given by
\begin{displaymath}
G_5^{-1}=\left(
\begin{array}{c}
D  \\
1+D 
\end{array}
\right) .
\end{displaymath}
Two inverse encoders are quite different, but the values of $m_v$ calculated from $G^{-1}G$ are equal, which results in almost the same MMSE.
\par

\section{Conclusion}
In this paper, we have shown that the soft-decision input to the main decoder in an SST Viterbi decoder is regarded as the innovation as well. Although this fact can be obtained from the definition of innovations for hard-decision data, this time we have discussed the subject from the viewpoint of mutual information and mean-square error. Then by combining the present result with that in~\cite{taji 19}, it has been confirmed that the input to the main decoder in an SST Viterbi decoder is regarded as the innovation. Moreover, we have obtained an important result. Note that the MMSE has been expressed in terms of the distribution of the encoded block for the main decoder in an SST Viterbi decoder. Then through the argument, the input-output mutual information has been connected with the distribution of the encoded block for the main decoder. We think this is an extension of the relation between the mutual information and the MMSE to coding theory.
\par
On the other hand, we have problems to be further discussed. Note that since the input is not Gaussian, the discussions are based on inequalities and approximate expressions. In particular, when a given QLI code is regarded as a general code, the difference between $\frac{\log(1+\rho)}{\rho}$ and $\frac{1}{2}\mbox{mmse}=\frac{1}{2}\xi(\alpha_1, \alpha_2, \delta)$ is not so small. We remark that this comparison has been done based on two inequalities regarding evaluation of the input-output mutual information. Moreover, the discussions are based partly on numerical calculations. Hence our argument seems to be slightly less rigorous in some places. Nevertheless, we think the closeness of two curves in Figs. 4 and 5 implies that the inference and the related results in this paper are reasonable.

% or


\appendices
\section{Proof of Lemma 4}
\renewcommand{\theequation}{\thesection.\arabic{equation}}
We use mathematical induction on $m$.
\par
1) $P(e_1=1)=\epsilon \leq \frac{1}{2}$ is obvious.
\par
2) Suppose that $P(e_1+\cdots +e_m=1)\leq \frac{1}{2}$. Let $q=P(e_1+\cdots +e_m=1)$. Then we have
\begin{eqnarray}
\lefteqn{P(e_1+\cdots +e_m+e_{m+1}=1)} \nonumber \\
&& =P(e_1+\cdots +e_m=0)P(e_{m+1}=1) \nonumber \\
&& +P(e_1+\cdots +e_m=1)P(e_{m+1}=0) \nonumber \\
&& =(1-q)\epsilon +q(1-\epsilon) \nonumber \\
&& =\epsilon+q(1-2\epsilon) .
\end{eqnarray}
Since $q \leq \frac{1}{2}$ by the assumption, we have
\begin{eqnarray}
P(e_1+\cdots +e_m+e_{m+1}=1) &\leq& \epsilon+\frac{1}{2}(1-2\epsilon) \nonumber \\
&=& \frac{1}{2} .
\end{eqnarray}

% end of file

\section{Proof of Lemma 5}
Suppose that $\alpha_1(\epsilon)$, $\alpha_2(\epsilon)$, and $\alpha_{11}(\epsilon)$ have been determined given $\epsilon$. Here note $\mbox{\boldmath $v$}_k=(v_k^{(1)}, v_k^{(2)})=(\mbox{\boldmath $e$}_kG^{-1})G$. Let $e$ be the (error) terms common to $v_k^{(1)}$ and $v_k^{(2)}$. Then $v_k^{(1)}$ and $v_k^{(2)}$ are expressed as
\begin{eqnarray}
v_k^{(1)} &=& e+\tilde e_1 \\
v_k^{(2)} &=& e+\tilde e_2 ,
\end{eqnarray}
where $e$, $\tilde e_1$, and $\tilde e_2$ are mutually independent. Set
\begin{equation}
\left\{
\begin{array}{l}
p=P(e=1) \\
s=P(\tilde e_1=1) \\
t=P(\tilde e_2=1) .
\end{array} \right.
\end{equation}
We have
\begin{eqnarray}
\alpha_{11} &=& (1-p)st+p(1-s)(1-t) \\
\alpha_1 &=& (1-p)s+p(1-s) \\
\alpha_2 &=& (1-p)t+p(1-t) .
\end{eqnarray}
By direct calculation, it is derived that
\begin{eqnarray}
\delta &=& \alpha_{11}-\alpha_1\alpha_2 \nonumber \\
&=& p(1-p)(1-2s)(1-2t) .
\end{eqnarray}
Since $0 \leq s \leq \frac{1}{2}$ and since $0 \leq t \leq \frac{1}{2}$ (see Lemma 4), it follows that $\delta \geq 0$.

% end of file

\section{Proof of Proposition 5}
Suppose that $\alpha_1(\epsilon)$, $\alpha_2(\epsilon)$, and $\alpha_{11}(\epsilon)$ have been determined given $\epsilon$. It suffices to show that
\begin{displaymath}
\alpha_1(1-\alpha_1)+\alpha_2(1-\alpha_2)-2\delta \geq 0 .
\end{displaymath}
We apply the same argument as that in the proof of Lemma 5. Let $p$, $s$, and $t$ be the same as those for Lemma 5. We have
\begin{eqnarray}
\alpha_1(1-\alpha_1) &=& p(1-p)+s(1-s)-4ps(1-p)(1-s) \\
\alpha_2(1-\alpha_2) &=& p(1-p)+t(1-t)-4pt(1-p)(1-t) \\
\delta &=& p(1-p)(1-2s)(1-2t) .
\end{eqnarray}
By direct calculation, it follows that
\begin{eqnarray}
\lefteqn{\alpha_1(1-\alpha_1)+\alpha_2(1-\alpha_2)-2\delta} \nonumber \\
&& =s(1-s)+t(1-t)+4p(1-p)(s-t)^2 \geq 0 .
\end{eqnarray}

% end of file

\section{Explanation for $\frac{d}{d \rho}I[x; \sqrt{\rho}\,x+w] \leq \frac{d}{d \rho}I[\tilde x; \sqrt{\rho}\,\tilde x+w]$}
The mutual information of Gaussian channel with equiprobable binary input has been given (see~\cite[p.1263]{guo 05}). Let $x$ be the equiprobable binary input with unit variance. Using the relation~\cite[Theorem 1]{guo 05}
\begin{equation}
\frac{d}{d\rho}I[x; \sqrt{\rho}\,x+w]=\frac{1}{2}\mbox{mmse}(\rho) ,
\end{equation}
we have
\begin{eqnarray}
\lefteqn{\frac{d}{d\rho}I[x; \sqrt{\rho}\,x+w]=\frac{1}{2}\mbox{mmse}(\rho)} \nonumber \\
&& =\frac{1}{2} \left(1-\int_{-\infty}^{\infty}\frac{e^{-\frac{y^2}{2}}}{\sqrt{2 \pi}}\tanh(\rho-\sqrt{\rho}\,y)dy \right) .
\end{eqnarray}
\par
On the other hand, for the standard Gaussian input $\tilde x$, we have
\begin{eqnarray}
\frac{d}{d\rho}I[\tilde x; \sqrt{\rho}\,\tilde x+w] &=& \frac{d}{d\rho}\left(\frac{1}{2}\log(1+\rho)\right) \nonumber \\
&=& \frac{1}{2}\frac{1}{1+\rho} .
\end{eqnarray}
\par
Accordingly, it suffices to show that
\begin{equation}
\frac{1}{1+\rho} \geq 1-\int_{-\infty}^{\infty}\frac{e^{-\frac{y^2}{2}}}{\sqrt{2 \pi}}\tanh(\rho-\sqrt{\rho}\,y)dy .
\end{equation}
Note that this inequality is equivalent to
\begin{equation}
\int_{-\infty}^{\infty}\frac{e^{-\frac{y^2}{2}}}{\sqrt{2 \pi}}\tanh(\rho-\sqrt{\rho}\,y)dy \geq \frac{\rho}{1+\rho} .
\end{equation}
Furthermore, the above is rewritten as
\begin{eqnarray}
\lefteqn{\int_0^{\infty}\frac{e^{-\frac{(y-\rho)^2}{2\rho}}}{\sqrt{2 \pi \rho}}\tanh(y)dy} \nonumber \\
&& -\int_0^{\infty}\frac{e^{-\frac{(y+\rho)^2}{2\rho}}}{\sqrt{2 \pi \rho}}\tanh(y)dy \nonumber \\
&& \geq \frac{\rho}{1+\rho} ,
\end{eqnarray}
where $\frac{1}{\sqrt{2 \pi \rho}}e^{-\frac{(y-\rho)^2}{2\rho}}$ represents the Gaussian distribution with mean $\rho$ and variance $\rho$.
\par
Here note the integrands on the left-hand side. The variable $\rho$ appears only in Gaussian distributions. Hence using a linear approximation of $\tanh(y)$, numerical integration is possible. $\tanh(y)~(0 \leq y < \infty)$ is approximated as
\begin{equation}
\tanh(y) \approx \left\{
\begin{array}{rl}
\displaystyle{y},& 0 \leq y \leq \frac{1}{2} \\
\displaystyle{\frac{1}{2}y+\frac{1}{4}},& \frac{1}{2} \leq y \leq 1 \\
\displaystyle{\frac{3}{10}y+\frac{9}{20}},& 1 \leq y \leq \frac{3}{2} \\
\displaystyle{\frac{1}{15}y+\frac{4}{5}},& \frac{3}{2} \leq y \leq 3 \\
\displaystyle{1},& 3 \leq y < \infty . \\ 
\end{array} \right.
\end{equation}
(cf. $\tanh(0)=0,~\tanh(0.5)\approx 0.4621,~\tanh(1.0)\approx 0.7616,~\tanh(1.5)\approx 0.9051,~\tanh(3.0)\approx 0.9951$)
\par
If the above inequality holds for each $\rho$, then $\frac{d}{d \rho}I[x; \sqrt{\rho}\,x+w] \leq \frac{d}{d \rho}I[\tilde x; \sqrt{\rho}\,\tilde x+w]$ will be shown. For example, let $\rho=1$. The target inequality becomes
\begin{eqnarray}
\lefteqn{\int_0^{\infty}\frac{e^{-\frac{(y-1)^2}{2}}}{\sqrt{2 \pi}}\tanh(y)dy} \nonumber \\
&& -\int_0^{\infty}\frac{e^{-\frac{(y+1)^2}{2}}}{\sqrt{2 \pi}}\tanh(y)dy \nonumber \\
&& \geq \frac{1}{2} .
\end{eqnarray}
By carrying out numerical integration, we have
\begin{eqnarray}
\lefteqn{\int_0^{\infty}\frac{e^{-\frac{(y-1)^2}{2}}}{\sqrt{2 \pi}}\tanh(y)dy} \nonumber \\
&& -\int_0^{\infty}\frac{e^{-\frac{(y+1)^2}{2}}}{\sqrt{2 \pi}}\tanh(y)dy \nonumber \\
&& \approx 0.6079-0.0665=0.5414 > \frac{1}{2} .
\end{eqnarray}
Hence we actually have $\frac{d}{d \rho}I[x; \sqrt{\rho}\,x+w] \leq \frac{d}{d \rho}I[\tilde x; \sqrt{\rho}\,\tilde x+w]$ at $\rho=1$.

% end of file

%\section{}

%\section*{Acknowledgment}

\ifCLASSOPTIONcaptionsoff
  \newpage
\fi

\end{document}